\documentstyle[12pt,twoside]{article}
\begin{document}

\newcommand{\beqn}{\begin{eqnarray}}
\newcommand{\eeqn}{\end{eqnarray}}
\newcommand{\be}{\begin{equation}}
\newcommand{\ee}{\end{equation}}
\newcommand{\ba}{\begin{array}}
\newcommand{\ea}{\end{array}}
\newcommand{\pa}{\partial}
\newcommand{\re}{\ref}
\newcommand{\ci}{\cite}
\newcommand{\la}{\label}
\newcommand{\fr}{\frac}
\newcommand{\ve}{\varepsilon}
\newcommand{\de}{\delta}
\newcommand{\al}{\alpha}
\newcommand{\ga}{\gamma}
\newcommand{\Ga}{\Gamma}
\newcommand{\si}{\sigma}
\newcommand{\ds}{\displaystyle}
\newcommand{\pr}{\prime}
\newcommand{\La}{\Lambda}
\newcommand{\De}{\Delta}
\newcommand{\Si}{\Sigma}
\newcommand{\ti}{\tilde}
\newcommand{\Om}{\Omega}
\newcommand{\om}{\omega}

\newcommand {\R}{{\rm\bf R}}
\newcommand{\T}{{\rm\bf T}}
\newcommand {\Z}{{\rm\bf Z}}
\newcommand {\N}{{\rm\bf N}}
\newcommand {\PV}{{\rm PV}}
\newcommand {\C}{{\rm\bf C}}
\renewcommand{\theequation}{\thesection.\arabic{equation}}

\newcommand{\const}{\mathop{\rm const}\nolimits}
\newcommand{\tr}{\mathop{\rm tr}\nolimits}
\newcommand{\tg}{\mathop{\rm tg}\nolimits}
\newcommand{\supp}{\mathop{\rm supp}\nolimits}
\newcommand{\sign}{\mathop{\rm sign}\nolimits}
\newcommand{\dist}{\mathop{\rm dist}\nolimits}
\newtheorem{theorem}{Theorem}[section]
\renewcommand{\thetheorem}{\arabic{section}.\arabic{theorem}}
\newtheorem{definition}[theorem]{Definition}
\newtheorem{lemma}[theorem]{Lemma}
\newtheorem{example}[theorem]{Example}
\newtheorem{remark}[theorem]{Remark}
\newtheorem{remarks}[theorem]{Remarks}
\newtheorem{cor}[theorem]{Corollary}
\newtheorem{pro}[theorem]{Proposition}

\newcommand{\bo}{{\hfill\loota}}
\newcommand{\loota}{\hbox{\enspace{\vrule height 7pt depth 0pt width 7pt}}}
%%%%%%%%%%%%%%%%%%%%%%%%%%%%%%%%%%%%%

\begin{titlepage}
\begin{center}
{\Large\bf Lattice Dynamics in the Half--Space, II. \\
\vspace{0.5cm}
Energy Transport Equation}\\
\vspace{2cm}
{\large T.V.~Dudnikova}
\footnote{Supported partly by
research grant of RFBR (09-01-00288)}\\
{\it Elektrostal Polytechnical Institute\\
 Elektrostal 144000, Russia}\\ 
e-mail:~tdudnikov@mail.ru
\end{center}
 \vspace{1cm}
 \begin{abstract}
We consider the lattice dynamics in the half-space. The initial
data are random according to a probability measure which enforces
slow spatial variation on the linear scale $\varepsilon^{-1}$. We
establish two time regimes. For times of order
$\varepsilon^{-\gamma}$, $0<\gamma<1$, locally the measure
converges to a Gaussian measure which is time stationary
with a covariance inherited from the initial measure 
(non-Gaussian, in general). For times of order $\varepsilon^{-1}$, 
this covariance changes in time and is governed by a
semiclassical transport equation.\bigskip
 \end{abstract}

\end{titlepage}
%%%%%%%%%%%%%%%%%%%%%%%%%%%%%%%%%%%%%
 \section{Introduction}
%%%%%%%%%%%%%%%%%%%%%%%%%%%%%%%%%%%

%The energy transport by oscillations of atoms in crystals 
%is a central problem of solid state physics.

The paper concerns a mathematical problem of foundations of
statistical physics and continues the work \cite{DS}
devoted to the derivation of %energy transport equation 
 a limiting "hydrodynamic" (Euler type) equation
from the Hamilton dynamics. 
We refer the reader to \cite{DeMasi, DSS, Sp91, Sp05}
for a detailed discussion of the results and methods
on this problem. 
%The present state of the problem, including new results
%and methods on this topic, is discussed in review papers 
%\cite{DeMasi, DSS, Sp91, Sp05} and Spohn's works \cite{Sp91,Sp05} 
%to which we refer the reader for the details.  

As the model we consider the harmonic crystals in the half-space 
$\Z^d_+=\{z\in\Z^d:z_1>0\}$.
In the harmonic approximation, the crystal is characterized
by the displacements $u(z,t)\in\R^n$,  $z\in\Z^d_+$, of the crystal atoms 
from their equilibrium positions. 
The field  $u(z,t)$ is governed by a discrete wave equation.

The derivation of hydrodynamic equations is connected
with the problem of convergence to an equilibrium measure. 
Hence, the first step in our inverstigation is the proof 
of such convergence. This step was done in \cite{D08}.  
We assume that a probability measure $\mu_0$ giving the distribution 
of initial data has some mixing properties. If $\mu_t$ denotes
the time-evolved measure at time $t$, then the limit
\be\la{0.1}
\lim_{t\to\infty}\mu_t=\mu_\infty
\ee
is established, where $\mu_\infty$ is an  equilibrium Gaussian measure.
(The precise formulation of this assertion is given by Theorem \ref{the1}).
In  \cite{DKS, DK2}, we have analyzed
the  long-time convergence to an equilibrium distribution for 
systems described by partial differential equations in $\R^d$.
In \ci{DKS1}--\cite{DS}, we extended the results to harmonic crystals. 
In the above-mentioned papers the systems were 
considered in the entire space. 
In \cite{D08}, the dynamics of the harmonic crystals 
  is studied first in the half-space $\Z^d_+$. 

To derive the hydrodynamic equation
we apply the special so-called hydrodynamic limit procedure.
%At first we change variables so that the spatial and temporal scales are
%of order of $\ve$.
Given a matrix function $\{R(r,\cdot),r\in\R^d\}$
(so-called "spectral density matrix function" in the terms of R.L.~Dobrushin
and others, \cite{DPST}) we consider a family of measures 
$\{\mu_0^\ve,\ve>0\}$ which satisfies the following conditions:
(i) For any $r\in\R^d_+$, the covariance $Q_\ve(z,z')$
of the measure $\mu_0^\ve$ at points $z,z'\in\Z^d_+$ close to the $[r/\ve]$
is approximately (as $\ve\to0$) described by $R(r,\cdot)$;
(ii) the covariance $Q_\ve(z,z')$ vanishes as $|z-z'|\to\infty$
uniformly in $\ve$ 
(see conditions {\bf V1} and {\bf V2} in Section 2.2 below).
Given nonzero $\tau\in\R$ and $r\in\R^d_+$, we study the distribution
$\mu_{\tau/\ve,r}^\ve$  of the random solution $u(z,t)$
at time moments $\tau/\ve$ and close to the spatial point $[r/\ve]$.
We establish the limit
\be\la{0.2}
\lim_{\ve\to0}\mu_{\tau/\ve,r}^\ve=\mu^G_{\tau,r},
\ee
where $\mu^G_{\tau,r}$ is a Gaussian measure (see Theorem \ref{the3}).
In particular, we derive the explicit formulas 
for covariance matrix $q^G_{\tau,r}(z-z')$
of the limit measure $\mu^G_{\tau,r}$. These formulas allow us 
to conclude that the matrix function $\hat q^G_{\tau,r}(\theta)$ 
evolves according to the following equation:
\beqn\la{0.3}
\partial_\tau f_{\tau,r}(\theta)=i\,C(\theta)\nabla\Omega(\theta)
\cdot \nabla_r f_{\tau,r}(\theta),\quad r\in\Z^d_+,\,\,\,\tau>0,
\eeqn 
where $C(\theta)=\left(\ba{cc}
0&\Omega^{-1}(\theta)\\
-\Omega(\theta)&0 \ea\right)$, and, roughly, $\Omega(\theta)$ is 
the dispersion relation of the harmonic
crystal. The boundary and initial conditions for (\ref{0.3})
are written in terms of the function $R(r,\cdot)$. 
The equation of type (\ref{0.3}) 
is called a hydrodynamic (or Euler) equation (see \cite{DPST}).  
The result (\ref{0.3}) 
is a continuation of the works \cite{DPST} and \cite{DS}. 
In \cite{DPST}, the problem has been studied for the infinite chain 
of harmonic oscillators on one-dimensional  lattice $\Z^1$. 
In \cite{DS}, the result has been extended to the many-dimensional case. 

In phonon physics it
is standard practice to use the Wigner function $W(t,r,\theta)$ as
density of phonons with wave number $\theta$ at location $r$ and
at specified time $t$. $W$ evolves according to the semiclassical
energy transport equation
 \be\la{I.1}
\partial_t W(t,r,\theta)= -\nabla\Omega(\theta)\cdot
 \nabla_r W(t,r,\theta)
\ee 
(see Theorem \ref{the2} and Corollary \ref{cor2.13}).
$W(t,r,\theta)$ at fixed $r,t$ are expressed by the
covariance $\hat q^G_{t,r}(\theta)$ which is invariant
under the lattice dynamics. 
Thus (\ref{0.3}) or (\ref{I.1}) can be understood as
the equations governing the motion of the parameters which
characterize the locally stationary measures. 
%%%%%%%%%%%%%%%%%%%%%%%%%%
\vspace{2mm}

{\bf Acknowledgment}

Author would like to thank Professor H. Spohn for the helpful discussions
 and the warm hospitality in the Munich Technical University. 

%%%%%%%%%%%%%%%%%%%%%%%%%%%%%%%
 \subsection{Model}
%%%%%%%%%%%%%%%%%%%%%%%%%%%%%%%%%%%

We study the dynamics of the harmonic crystals 
in $\Z^d_+$, $d\ge 1$, 
\beqn\la{1+}
\ddot u(z,t)=-\sum\limits_{z'\in \Z^d_+}\left(V(z-z')-V(z-\tilde z')\right)
u(z',t),\,\,\,\,z\in\Z^d_+,\,\,\,\,t\in\R,
\eeqn
with zero boundary condition,
\be\la{2+}
u(z,t)|_{z_1=0}=0,
\ee
and with the initial data
\be\la{3+}
u(z,0)=u_0(z),\quad \dot u(z,0)=u_1(z),\quad z\in\Z^d_+.
\ee
Here $\Z^d_+=\{z\in \Z^d:\,z_1>0\}$, $\tilde z=(-z_1,z_2,\dots, z_d)$,
$V(z)$ is the interaction (or force) matrix, $\left(V_{kl}(z)\right)$,
$k,l=1,\dots,n$,  $u(z,t)=(u_1(z,t),\dots,u_n(z,t))$,
$u_0(z)=(u_{01}(z),\dots,u_{0n}(z))\in\R^n$, and
correspondingly for $u_1(z)$. 
To coordinate the boundary and initial conditions, we assume
that  $u_0(z)=u_1(z)=0$ for $z_1=0$.

Write $Y(t)=(Y^0(t),Y^1(t))\equiv (u(\cdot,t),\dot u(\cdot,t))$ and
$Y_0=(Y_0^0,Y_0^1)\equiv (u_0(\cdot),u_1(\cdot))$. Then
(\ref{1+})--(\ref{3+}) becomes the evolution equation
\be\label{CP1}
\dot Y(t)={\cal A}_+Y(t),\quad t\in\R,\,\,z\in\Z^d_+,
\quad Y^0(t)|_{z_1=0}=0,\quad Y(0)=Y_0.
\ee
Here ${\cal A}_+=\left(\ba{cc}0&1\\-{\cal V}_+&0\ea\right)$
with 
${\cal V}_+u(z):= \sum\limits_{z'\in\Z^d_+}(V(z-z')-V(z-\tilde z'))u(z')$.

Let us assume that
\be\label{condE0}
V(z)=V(\tilde z),\quad \mbox{where }\,\tilde z=(-z_1,\bar z),
\quad\bar z=(z_2,\dots,z_d)\in \Z^{d-1}.
\ee
Then the solution to the problem  (\ref{CP1}) can be represented as the 
restriction of the solution to the Cauchy problem
with  odd initial data on the half-space. 
More precisely, consider the following Cauchy problem for the harmonic crystal
in the entire space $\Z^d$:
\beqn\la{CP1''}
\left\{\ba{l}
\ddot v(z,t)=-\sum\limits_{z'\in \Z^d}V(z-z')v(z',t),\,\,\,\,z\in\Z^d,
\,\,\,\,t\in\R,\\
v(z,0)=v_0(z),\quad \dot v(z,0)=v_1(z),\quad z\in\Z^d.
\ea\right.
\eeqn
%%%%%%%%%%%%%%%%%%%%%%%%%%%%%%%%%%%%%
%Formally, this is a linear Hamiltonian system
%with the Hamiltonian functional
% \be\la{H} H(Y)= \frac{1}{2} \langle v,v\rangle
%+\frac{1}{2} \langle  {\cal V}u, u\rangle, \quad Y=(u,v), \ee
%where ${\cal V}$ is the convolution
%operator with the matrix kernel $V$, the kinetic energy is given by
%$\ds\frac{1}{2} \langle v,v\rangle=
%\frac{1}{2} \sum_{x\in\Z^d}|v(x)|^2$, and the potential
%energy by $\ds\frac{1}{2}\langle{\cal V}  u,u\rangle
%=\frac{1}{2}\sum_{x,y\in\Z^d} u(x)\cdot V(x-y)u(y)$. Here
%``$\cdot$'' stands for the  scalar product in the
%Euclidean space $\R^n\!$, resp. in $\R^d\!$.
%%%%%%%%%%%%%%%%%%%%%%%%%%%%%%%%%%%
Write $X(t)=(X^0(t),X^1(t))\equiv (v(\cdot,t),\dot v(\cdot,t))$ and
$X_0=(X_0^0,X_0^1)\equiv (v_0(\cdot),v_1(\cdot))$. Then
(\ref{CP1''}) becomes
\be\la{CP1'}
\dot X(t)={\cal A}X(t),\quad t\in\R,\quad X(0)=X_0.
\ee
Here ${\cal A}=\left(\ba{cc}0&1\\-{\cal V}&0\ea\right)$,
where ${\cal V}$ is a convolution operator with the matrix kernel $V$.

Assume that the initial data $X_0(z)$ form an odd function with respect to
 $z_1\in\Z^1$, i.e., let $X_0(z)=-X_0(\tilde z)$.
Then the solution $v(z,t)$ of (\ref{CP1''})
is also an odd function with respect to $z_1\in\Z^1$.
Restrict the solution $v(z,t)$ 
to the domain $\Z^d_+$ and set $u(z,t)=v(z,t)|_{z_1\ge0}$.
Then $u(z,t)$ is the solution to the problem (\ref{1+}) with
the initial data $Y_0(z)=X_0(z)|_{z_1\ge0}$. 
\smallskip

  Assume that the initial data $Y_0$ for (\ref{CP1})
belong to the phase space ${\cal H}_{\al,+}$, $\al\in\R$, defined below.
%%--------------------------------
 \begin{definition}                 \la{d1.1}
 $ {\cal H}_{\al,+}$ is the Hilbert space
of $\R^n\times\R^n$-valued functions  of $z\in\Z^d_+$
 endowed  with  the norm
 \beqn\nonumber 
 \Vert Y\Vert^2_{\al,+}
 = \sum_{z\in\Z^d_+}\vert Y(z)\vert^2(1+|z|^2)^{\al} <\infty.
 \eeqn  
\end{definition}
%%%%%%%%%%%%%%%%%%%%%%%%%%%%%%%%%%%%%%

In addition, it is assumed that the initial data vanish ($Y_0=0$)
 at $z_1=0$.\smallskip

We impose the following conditions {\bf E1}--{\bf E6} on the matrix $V$.
\medskip\\
{\bf E1}. There are positive constants $C$ and $\gamma$ such that
$\|V(z)\|\le C e^{-\gamma|z|}$ for $z\in \Z^d$, where
$\|V(z)\|$ stands for the matrix norm.
\medskip

Let $\hat V(\theta)$ be the Fourier transform of $V(z)$ with the
convention
 $$
 \hat V(\theta)=
\sum\limits_{z\in\Z^d}V(z)e^{iz\cdot\theta}\,,\,\,\,\,\theta \in \T^d,
 $$
where "$\cdot$" stands for the inner product in Euclidean space $\R^d$
and  $\T^d$ for the $d$-torus $\R^d/(2\pi \Z)^d$.
\medskip\\
{\bf E2}. $ V$ is real and symmetric, i.e., $V_{lk}(-z)=V_{kl}(z)\in \R$,
$k,l=1,\dots,n$, $z\in \Z^d$.
\medskip

The two conditions imply that $\hat V(\theta)$ is a real-analytic
 Hermitian matrix-valued function of $\theta\in \T^d\!$.
\medskip\\
{\bf E3}. The matrix $\hat V(\theta)$ is  non-negative definite for
every $\theta \in \T^d$.
\medskip

Let us define the Hermitian  non-negative definite matrix,
 \be\label{Omega}
 \Omega(\theta)=\big(\hat V(\theta )\big)^{1/2}\ge 0.
 \ee
The matrix $\Omega(\theta)$  has the eigenvalues 
$0\leq\omega_1(\theta)<\omega_2(\theta) \ldots <\omega_s(\theta)$, 
$s\leq n$, and the
corresponding spectral projections $\Pi_\sigma(\theta)$ with
multiplicity $r_\sigma=\tr\Pi_\sigma(\theta)$. The mapping 
$\theta\mapsto\omega_\sigma(\theta)$ is the $\sigma\!$-th band function.
There are special points in $\T^d$ at which
the bands cross, which means that $s$ and $r_\sigma$ jump to
some other value. Away from such crossing points, $s$ and
$r_\sigma$ are independent of $\theta$. 
More precisely, the following lemma holds.
%%%%%%%%%%%%%%%%%%%%%%%%%%%%%%%%%%%%%%%%%%%%
\begin{lemma}\label{lc*} (see \cite[Lemma 2.2]{DKS1}).
Let conditions {\bf E1} and {\bf E2} hold. 
Then there exists a closed subset ${\cal C}_*\subset \T^d$
such that the following assertions hold:\\
(i) the Lebesgue measure of ${\cal C}_*$ is zero;\\
(ii) for any point $\Theta\in \T^d\setminus{\cal C}_*$, 
there exists a neighborhood ${\cal O}(\Theta)$ such that each band
function $\omega_\sigma(\theta)$ can be chosen 
as a real-analytic function on ${\cal O}(\Theta)$;\\
(iii) the eigenvalue $\omega_\sigma(\theta)$ has constant multiplicity
in $\T^d\setminus{\cal C}_*$;\\
(iv) the following spectral decomposition holds: 
 \be\label{spd'}
\Omega(\theta)=\sum_{\sigma=1}^s \omega_\sigma
(\theta)\Pi_\sigma(\theta),\quad \theta\in \T^d\setminus{\cal C}_*, 
 \ee
 where $\Pi_\sigma(\theta)$ is an orthogonal projection in
$\R^n$, and $\Pi_\sigma$ is a real-analytic function on
$\T^d\setminus{\cal C}_*$.
\end{lemma}
%%%%%%%%%%%%%%%%%%%%%%%%%%%%%%%%%%%%%%%

For $\theta\in \T^d\setminus{\cal C}_*$, denote by
Hess$(\omega_\sigma)$ the matrix of second partial derivatives. 
The next condition on $V$ is as follows.
\smallskip\\
{\bf E4}. Let $D_\sigma(\theta)
=\det\big(\rm{Hess}(\omega_\sigma(\theta))\big)$.
Then $D_\sigma(\theta)$ does not vanish identically on
$\T^d\setminus{\cal C}_*$, $\sigma=1,\ldots,s$.
\medskip

Let us write
 \be\label{c0ck}
{\cal C}_0=\{\theta\in \T^d:\det \hat
 V(\theta)=0\}\,\, \mbox{and }\,
{\cal C}_\sigma=\{\theta\in \T^d\setminus {\cal
C}_*:\,D_\sigma(\theta)=0\},\,\,\, \sigma=1,\dots,s.
 \ee
Then the Lebesgue measure of ${\cal C}_\sigma$ vanishes, $\sigma=0,1,...,s$
(see \cite[Lemma 2.3]{DKS1}).

The last two conditions on $V$ look as follows.
\medskip\\
{\bf E5}.  For each $\sigma\ne \sigma'$, 
the identities $\omega_\sigma(\theta)
\pm\omega_{\sigma'}(\theta)\equiv\const_\pm$ for 
$\theta\in \T^d\setminus {\cal C}_*$, do not hold  with $\const_\pm\ne 0$.
\medskip

This condition holds trivially  for $n=1$.\medskip\\
{\bf E6}. $\Vert \hat V^{-1}(\theta)\Vert\in L^1(\T^d)$.
\medskip

If ${\cal C}_0=\emptyset$, then $\|\hat{V}^{-1}(\theta)\|$ is
bounded, and {\bf E6} holds trivially.
%%%%%%%%%%%%%%%%%%%%%%%%%%%%%%%%%%%%%%%%%%%%%%%%%%
\begin{remark}
{\rm Conditions {\bf E1}--{\bf E6} are satisfied, in particular,
in the case of the {\it nearest neighbor crystal} in which 
the interaction matrix $V(z)=(V_{kl}(z))_{k,l=1}^n$ is of the form
$$
V_{kl}(z)=0\,\,\mbox{ for }\,k\not=l,\,\,\,\,
V_{kk}(z)=\left\{\ba{ll}-\gamma_k&\mbox{for }\, |z|=1,\\
2\gamma_k+m_k^2& \mbox{for }\, z=0,\\
0&\mbox{for }\, |z|\ge2,\ea\right.  \quad k=1,\dots,n,
$$ 
with $\gamma_k>0$ and $m_k\ge0$.
In this case, equation (\ref{1+}) becomes
$$
\ddot u_k(z,t)=(\gamma_k\Delta_L-m_k^2)u_k(z,t),\quad
k=1,\dots,n.
$$
Here $\Delta_L$ stands for the discrete
Laplace operator on the lattice $\Z^d$,
$$
\Delta_L u(z):= \sum\limits_{e,|e|=1}(u(z+e)-u(z)).
$$
Therefore, the eigenvalues of  $\hat V(\theta)$ are
 \be\la{omega}
 \tilde{\omega}_k(\theta)=
\sqrt{\, 2 \gamma_1(1-\cos\theta_1)+...+2 \gamma_d
(1-\cos\theta_d)+m_k^2}\,,\quad k=1,\dots,n.
 \ee
These eigenvalues still have to be labelled according to magnitude
and degeneracy as in Lemma \ref{lc*}.
Clearly, conditions {\bf {E1}}--{\bf {E5}} hold with ${\cal C}_*=\emptyset$.
If $m_k>0$ for any $k$, then the set ${\cal C}_0$ is empty and
condition {\bf E6} holds automatically. Otherwise, if $m_k=0$ for
some $k$, then ${\cal C}_0=\{0\}$. In this case,
{\bf E6} is equivalent to the condition $\om_k^{-2}(\theta)\in
L^1(\T^d)$, which holds if $d\ge 3$. Therefore, 
conditions {\bf E1}--{\bf E6} hold  if either (i) $d\ge 3$
or (ii) $d=1,2$ and $m_k >0$ for any $k$.}
\end{remark}
%%%%%%%%%%%%%%%%%%%%%%%%%%%%%%%%%%%%%%%%%
%\begin{pro} \la{p1.1} (see \ci[Proposition 2.5]{DKS1}).
%Let conditions {\bf E1} and {\bf E2} hold and choose some $\al\in\R$. Then \\
%(i) for any  $X_0 \in {\cal H}_\al$, there exists  a unique solution
%$(v(\cdot,t),\dot v(\cdot,t))\in C(\R, {\cal H}_\al)$
% to the Cauchy problem (\re{CP1'}).\\
%(ii) The operator $U(t):X_0\mapsto X(t)$ is continuous in ${\cal H}_\al$.
%\end{pro}
%%----------------------
\begin{lemma}\label{c1}(see \cite[Corollary 2.4]{D08})
Let conditions {\bf E1} and {\bf E2} hold. 
Choose some $\al\in\R$. Then
(i) for any  $Y_0 \in {\cal H}_{\al,+}$, there exists  a unique solution
$Y(t)\in C(\R, {\cal H}_{\al,+})$  to the mixed problem (\re{CP1});\\
(ii) the operator  $U_+(t):Y_0\mapsto Y(t)$ is continuous
on ${\cal H}_{\al,+}$.
\end{lemma}
%%----------------------------

The proof is based on the following formula for 
the solution  $X(t)$ of (\ref{CP1'}):  
\be\la{solGr}
X(t)=\sum\limits_{z'\in\Z^d}{\cal G}_t(z-z')X_0(z'),
\ee
where the function ${\cal G}_t(z)$ has the Fourier representation   
\be\la{Grcs}  
{\cal G}_t(z):= F^{-1}_{\theta\to z}[  
\exp\big(\hat{\cal A}(\theta)t\big)]  
=(2\pi)^{-d}\int\limits_{\T^d}e^{-iz\cdot \theta}  
\exp\big(\hat{\cal A}(\theta)t\big)\,d\theta  
\ee  
with 
\be\la{hA}   
\hat{\cal A}(\theta)=\left( \begin{array}{cc}   
0 & 1\\   
-\hat V(\theta) & 0   
\end{array}\right),\,\,\,\,\theta\in \T^d.    
\ee   
Therefore, 
the solution  $Y(t)$ of (\ref{CP1}) admits the representation   
\beqn\la{sol}
Y(t)=\sum\limits_{z'\in\Z^d_+} {\cal G}_{t,+}(z,z')
Y_0(z'),\quad z\in\Z^d_+,\\
\mbox{where }\,\,\,
{\cal G}_{t,+}(z,z'):={\cal G}_t(z-z')-{\cal G}_t(z-\tilde z').
\la{sol1}
\eeqn

%%%%%%%%%%%%%%%%%%%%%%%%%%%%%%%%%%%%%%%%%%%%%%%%%%
%%%%%%%%%%%%%%%%   section 2 %%%%%%%%%%%%%%%%%%
%%%%%%%%%%%%%%%%%%%%%%%%%%%%%%%%%%%%%%%%%%%%%%%%%%
\setcounter{equation}{0}
 \section{Main results}
%%-----------------------------------
%%%%%%%%%%%%%%%%%%%%%%  2.1    %%%%%%%%%%%%%%%%%%%%%%
\subsection{Convergence to equilibrium}\la{sec2.1}
%"Space-homogeneous" random initial data
%%%%%%%%%%%%%%%%%%%%%%%%%%%%%%%%%%%%%%%%%%%%

Denote by $\mu_0$ a Borel probability measure   
on ${\cal H}_{\alpha,+}$ giving the distribution of $Y_0$.
The expectation with respect to $\mu_0$  is denoted by $E_0$.

Assume that the initial measure $\mu_0$  
has the following properties {\bf S1}--{\bf S4}.
\smallskip\\
{\bf S1}. $Y_0(z)$ has zero expectation value,
$E_0 \big(Y_0(z)\big) = 0$ for $z\in\Z^d_+$.
\smallskip

For $a,b,c \in \C^n$,  denote by $a\otimes b$ the
 linear operator $(a\otimes b)c=a\sum^n_{j=1}b_j c_j$.
Denote by $ {\cal H}_\al$ the Hilbert space
of $\R^n\times \R^n$-valued functions  of $z\in\Z^d$
 endowed  with  the norm
 \beqn  \nonumber
 \Vert X\Vert^2_{\al} = \sum_{z\in\Z^d}\vert X(z)\vert^2
(1+|z|^2)^{\al} <\infty.
\eeqn
%%---------------------- 
\begin{definition}
A measure $\nu$ is said to be translation invariant if 
$\nu(T_h B)= \nu(B)$ for $B\in{\cal B}({\cal H}_{\alpha})$
and $h\in\Z^d$, where $T_h X(z)= X(z-h)$ for $z\in\Z^d$.   
\end{definition}
%%-----------------------------------------
{\bf S2}. The correlation matrices of the measure $\mu_0$ have the form
 \beqn \label{1.9'}
Q^{ij}_0(z,z')= E_0\big(Y^i_0(z)\otimes {Y^j_0(z')}\big)
=q^{ij}_0(z_1,z'_1,\bar z-\bar z'),\,\,\,z,z'\in\Z^d_+,\,\,\,i,j=0,1,
 \eeqn
where 
(i) $q^{ij}_0(z_1,z'_1,\bar z)=0$ for $z_1=0$ or $z_1'=0$,\\
(ii) $\lim_{y\to+\infty} q^{ij}_0(z_1+y,y,\bar z)
={\bf q}_0^{ij}(z)$, $z=(z_1,\bar z)\in\Z^d$.
Here ${\bf q}_0^{ij}(z)$ are correlation functions of some 
translation invariant measure $\nu_0$ with zero mean value 
on ${\cal H}_{\alpha}$.\\
%%----------------------------------
{\bf S3}. The measure $\mu_0$  has finite variance 
and finite mean energy density,
 \beqn\label{med}
e_0(z)=E_0 \big(\vert Y_0^0(z)\vert^2
 + \vert Y_0^1(z)\vert^2\big)
=\tr\left[Q_0^{00}(z,z)+Q_0^{11}(z,z)\right]\le e_0<\infty,\,\,\,z\in\Z^d_+.
 \eeqn

Finally, it is assumed that the measure $\mu_0$ satisfies a mixing condition.
To formulate this condition, denote by $\sigma ({\cal A})$,
${\cal A}\subset \Z^d_+$, the $\sigma $-algebra on
${\cal H}_{\alpha,+}$ generated by $Y_0(z)$ with $z\in{\cal A}$.
Define the Ibragimov mixing coefficient of the probability  measure
$\mu_0$ on ${\cal H}_{\alpha,+}$ by the rule (cf. \cite[Definition 17.2.2]{IL})
 \beqn \label{ilc} 
\varphi(r)= \sup_{\scriptsize{\ba{cc} {\cal A},{\cal B}\subset \Z^d_+\\
\dist({\cal A},\,{\cal B})\geq r \ea}}
\sup_{\scriptsize{
\ba{cc} A\in\sigma({\cal A}),B\in\sigma({\cal B})\\ \mu_0(B)>0\ea}}
\frac{|\mu_0(A\cap B) - \mu_0(A)\mu_0(B)|}{ \mu_0(B)}.
 \eeqn
%-----------------------------------
\begin{definition}
 A measure $\mu_0$ is said to satisfy 
the strong uniform Ibragimov mixing condition if
$\varphi(r)\to 0$ as $r\to\infty$.
\end{definition}
%---------------------------------------
{\bf S4}. The measure $\mu_0$ satisfies the strong uniform
Ibragimov mixing condition with
 \be\label{1.12}
\int\limits_{0}^{\infty}
 r^{d-1}\varphi^{1/2}(r)\,dr <\infty\,.
 \ee

%%%%%%%%%%%%%%%%%%%%%%%%%%%%%
\begin{remark}
{\rm The {\it uniform Rosenblatt mixing condition} \cite{Ros} 
is also sufficient, 
together with a higher power $>2$  in the bound (\ref{med}).
Namely, there is a $\delta >0$ such that
$$
E_0 \Big( \vert Y^0_0(z)\vert^{2+\delta}+\vert Y^1_0(z)\vert^{2+\delta}
\Big) \le C <\infty,\,\,\,\,  z\in\Z^d_+.
$$
Condition (\ref{1.12}) needs a modification, namely,
$
\ds\int\limits_0^{+\infty}\ds r^{d-1}\alpha^{p}(r)dr <\infty$ with 
$p=\min(\delta/(2+\delta), 1/2)$.
Here $\alpha(r)$ is the Rosenblatt mixing coefficient  defined
as in  (\ref{ilc}) but without $\mu_0(B)$ in the denominator.
Under these modifications, the statements of Theorem \ref{the1} 
and their proofs remain essentially unchanged.

 The  uniform Rosenblatt mixing condition can also be weakened,
see Remarks 3.4 in \cite{D08}. }
\end{remark} 
%%%-----------------------------------
\begin{definition}  \label{dmut}
We define $\mu_t$ as the Borel probability measure on ${\cal H}_{\alpha,+}$
 which gives the distribution of the random solution $Y(t)$,
\beqn\nonumber
\mu_t(B) =
\mu_0(U_+(-t)B),\,\,\,\mbox{where }\, B\in {\cal B}({\cal H}_{\alpha,+})
\,\,\,\mbox{and }\,t\in \R\,.
 \eeqn
%(ii) The correlation functions of the  measure $\mu_t$ are  defined by
%\be\label{qd}
%Q_t^{ij}(z,z')= E \Big(Y^i(z,t)\otimes
% Y^j(z',t)\Big),\,\,\,i,j= 0,1,\,\,\,\,z,z'\in\Z^d_+.\ee
%Here $Y^i(z,t)$ are the components of the random solution
%$Y(t)=(Y^0(\cdot,t),Y^1(\cdot,t))$ to the problem (\ref{CP1}).
\end{definition}
%%---------------

In \ci{D08}, we prove the weak convergence of the measures $\mu_t$
 on the space ${\cal H}_{\alpha,+}$ 
with $\al<-d/2$  to a limit measure $\mu_\infty$,
\be\label{1.8i}
\mu_t \,\buildrel {\hspace{2mm}{\cal H}_{\alpha,+}}\over
{- \hspace{-2mm} \rightharpoondown }
\mu_\infty\quad{\rm as}\,\,\,\, t\to \infty,
\ee
where $\mu_\infty$ is an equilibrium Gaussian measure on 
${\cal H}_{\alpha,+}$.  This means the convergence
$$
 \lim_{t\to\infty}\int f(Y)\,\mu_t(dY)=\int f(Y)\,\mu_\infty(dY)
$$
for any bounded continuous functional $f$ on ${\cal H}_{\alpha,+}$.
%%------------------------------------------------------
\begin{theorem} \label{the1}(see \cite{D08}).
 Let $d,n\ge 1$, $\alpha<-d/2$. Assume that conditions 
(\ref{condE0}), {\bf E1}--{\bf E6}, and {\bf S1}--{\bf S4} hold. Then\\
(i) the convergence in (\ref{1.8i}) holds. \smallskip\\
(ii) The limit measure $ \mu_\infty$ is a Gaussian
measure on ${\cal H}_{\alpha,+}$.\smallskip\\
(iii) The correlation matrices of the measures $\mu_t$
converge to a limit for $i,j=0,1$,
\be\label{corf}
Q^{ij}_t(z,z')=\int\big( Y^i(z)\otimes Y^j(z')\big)
\,\mu_t(dY)\to Q^{ij}_\infty(z,z'),\,\,\,\,t\to\infty,\quad
z,z'\in\Z^d_+.
\ee
The correlation matrix $Q_\infty(z,z')=(Q^{ij}_\infty(z,z'))_{i,j=0}^1$
of the limit measure $\mu_{\infty}$ has the form
\be\label{1.13}
Q_\infty(z,z')=q_\infty(z-z')-q_\infty(z-\tilde z')-q_\infty(\tilde z-z')+
q_\infty(\tilde z-\tilde z'),\quad z,z'\in\Z^d_+.
\ee
Here $q_\infty(z)=  q^+_{\infty}(z)+ q^-_{\infty}(z)$,
where in the Fourier transform, we have
\beqn
 \hat q^+_{\infty}(\theta)&=&\frac{1}{4} \sum\limits_{\sigma=1}^s
\Pi_\sigma(\theta)\left(\hat {\bf q}_0(\theta) 
+C(\theta)\hat {\bf q}_0(\theta)C(\theta)^*\right)\Pi_\sigma(\theta),
\label{1.14}\\
\hat q^-_{\infty}(\theta)&=&\frac{i}{4}\sum\limits_{\sigma=1}^s
{\rm sign} \left(\partial_{\theta_1}\omega_\sigma(\theta)\right)
 \Pi_\sigma(\theta) 
 \left(C(\theta)\hat {\bf q}_0(\theta) -
\hat {\bf q}_0(\theta)C(\theta)^*\right)\Pi_\sigma(\theta),
\,\,\,\theta\in\T^d\setminus {\cal C}_*,\,\label{1.15}
\eeqn 
$\Pi_\sigma(\theta)$ is the spectral projection in Lemma \ref{lc*} (iv), 
and  
\be\label{C(theta)}
C(\theta)=\left(\ba{cc}
0&\Omega(\theta)^{-1}\\
-\Omega(\theta)&0 \ea\right)\,,\quad 
C(\theta)^*=\left(\ba{cc} 0&-\Omega(\theta)\\
\Omega(\theta)^{-1}&0 \ea\right).
 \ee
(iv) The measure $\mu_\infty$ is time stationary, i.e., 
$[U_+(t)]^*\mu_\infty=\mu_\infty$, $t\in\R$.
\end{theorem}
%%---------------------
\begin{remark}
{\rm (i) From formulas (\ref{1.14})--(\ref{C(theta)}) it follows that
$\hat q_{\infty}(\theta)$ satisfies the "equilibrium condition", i.e.,
one has the form
\be\la{2.10'}
\hat q_{\infty}(\theta)=\left(
\ba{cc} h(\theta)&g(\theta)\\
-g(\theta)&\Omega^2(\theta)h(\theta)
\ea\right).
\ee
Moreover, $(\hat q^{ij}_{\infty}(\theta))^*=\hat q^{ji}_{\infty}(\theta)$,
$i,j=0,1$, and 
$\hat q^{00}_{\infty}(\theta)\ge0$, $\hat q^{11}_{\infty}(\theta)\ge0$.
\smallskip

(ii) Introduce the complex-valued field
 \beqn\la{2.1}
 a(x)= \frac{1}{\sqrt{2}}\Big({\cal V}_+^{1/4}
 u_0(x)+i{\cal V}_+^{-1/4}u_1(x)\Big)\in \C^n\,,
\quad x\in \Z^d\,,
 \eeqn
where 
$$
{\cal V}_+^{k}u:=\sum\limits_{z\in\Z^d_+}(V^k(x-z)-V^k(x-\ti z))u(z),\quad
\mbox{with }\,V^k(z):=F^{-1}_{\theta\to z}\left(\hat V^{k}(\theta)\right).
$$ 
Let $a(x)^*$ stand for complex conjugate field.
Obviously $E_t \big(a(x)\big)=0$. The covariance $Q_\infty(x,y)$ has two
parts. By Theorem \ref{the1}, the $aa$-, equivalently the $a^\ast a^\ast$-, 
covariance satisfies
 \beqn\nonumber
 \lim_{t\to\infty}E_t \big(a(x)\otimes a(y)\big)=0\,.
 \eeqn
For the $a^\ast a$-covariance, Theorem \ref{the1} (iii) implies
 \beqn\nonumber%\la{2.3}
\lim_{t\to\infty} E_t \big(a(x)^\ast\otimes a(y)\big)=
  W(x-y)-W(x-\ti y)-W(\ti x-y)+W(\ti x-\ti y),\quad x,y\in\Z^d_+,
 \eeqn
where in Fourier transform
\beqn\nonumber
\hat W(\theta)=
%\frac12\left(\sqrt\Omega(\theta)\,\hat q^{00}_\infty(\theta)\sqrt\Omega(\theta)
%+\frac{1}{\sqrt\Omega(\theta)}\hat q^{11}_\infty(\theta)
%\frac{1}{\sqrt\Omega(\theta)}
%+i\sqrt\Omega(\theta)\,\hat q^{01}_\infty(\theta)\frac{1}{\sqrt\Omega(\theta)}
%\right.\nonumber\\
%&&\left.-i\frac{1}{\sqrt\Omega(\theta)}\,\hat q^{10}_\infty(\theta)\sqrt\Omega(\theta)
%\right)=
\Omega(\theta)\, \hat q^{00}_\infty(\theta)+i\,\hat q^{01}_\infty(\theta)
= \frac12\sum\limits_{\sigma=1}^s
\Big(1-\sign (\nabla_{\theta_1}\omega_\sigma(\theta))\Big)
\Pi_\sigma(\theta)\hat W_0(\theta)\Pi_\sigma(\theta),\nonumber
\eeqn
with 
$
\hat W_0(\theta)=\frac12\left[\omega_\sigma(\theta)\hat q^{00}_0(\theta)
+\omega_\sigma^{-1}(\theta)\hat q^{11}_0(\theta)
+i\,\hat q^{01}_0(\theta)-i\,\hat q^{10}_0(\theta)\right].
$ 
%\beqn
%\hat W(\theta)&=&\hat W^+(\theta)+\hat W^-(\theta),\nonumber\\
%\hat W^+(\theta)&=&S_-\hat q^+_\infty S_+^T
%=\frac12\sum\limits_{\sigma=1}^s\Pi_\sigma S_-\hat q_0S_+^T\Pi_\sigma,\nonumber\\
%\hat W^-(\theta)&=&S_-\hat q^+_\infty S_+^T
%=\frac12\sum\limits_{\sigma=1}^s
%\sign (\partial_{\theta_1}\omega_\sigma(\theta))
%\Pi_\sigma S_-\hat q_0S_+^T\Pi_\sigma,\nonumber
%\eeqn
%where 
%$$S_-\hat q_0S_+^T=\Omega\, \hat q^{00}_0+i\,\hat q^{01}_0.$$
}
\end{remark}
%%%%%%%%%%%%%%%%%%%%%%%%%%%%%
%{\bf Example:} {\it nearest neighbor crystal}.
%$$\hat W(\theta)_{kl}=\frac14\left(1-\sign(\sin\theta_1)\right)
%\sum\limits_{\sigma=1}^s\chi_{kl}(\sigma)\hat W_0(\theta)_{kl},
%\quad k,l=1,\dots,n,$$
%where
%$$\chi_{kl}(\sigma)=\left\{\ba{cl}
%1&\mbox{if }\,k,l\in[r_{\sigma-1}+1,r_\sigma]\\
%0&\mbox{otherwise},
%\ea\right.\quad \sigma=1,\dots,s.$$

%%%%%%%%%%%%%%%%%%   2.2 %%%%%%%%%%%%%%%%%%
\subsection{Initial measure with slow variation}\la{sec.2}
%%%%%%%%%%%%%%%%%%%%%%%%%

Let $\{\mu^\varepsilon_0,\varepsilon>0\}$ be a family of initial
measures. Roughly, in a linear region of size $\varepsilon^{-1}$,
$\varepsilon\ll 1$, $\mu^\varepsilon_0$ looks like the 
initial measure from Section \ref{sec2.1}. 
To formulate the main conditions {\bf V1}--{\bf V2} on the initial covariance,
let us introduce  the complex $2n\times 2n$
matrix-valued function $R(r,x,y)=(R^{ij}(r,x,y))^1_{i,j=0}$, $r\in\R^d$,
 $x,y\in\Z^d_+$, with the following properties.
\medskip\\
{\bf I0}. $R(r,x,y)=0$ for $x_1=0$ or $y_1=0$.  
The $n\times n$ matrix-valued functions $R^{ij}(r,x,y)$
have the form
$$
R^{ij}(r,x,y)={\bf R}^{ij}(r,x_1,y_1,\bar x-\bar y),\quad\mbox{where }\,
x=(x_1,\bar x),\,\,y=(y_1,\bar y),\,\,i,j=0,1.
$$
Moreover,
\be\la{3.14}
\lim_{y_1\to+\infty} {\bf R}^{ij}(r,y_1+z_1,y_1,\bar z)={\bf R}_0^{ij}(r,z), 
\quad z=(z_1,\bar z)\in\Z^d,\quad i,j=0,1.
\ee
%%----------------------
{\bf I1}. For every fixed $r\in\R^d$ and $i,j=0,1$, 
the bound holds,
\be\la{3.2} |R^{ij}(r,x,y)|\le C(1+|x-y|)^{-\gamma},\,\,\,\,x,y\in\Z^d_+,
 \ee
where $C$ is some  positive constant, $\gamma >d$.
In particular, for every $r\in\R^d$,
\be\la{3.2'} 
|{\bf R}_0^{ij}(r,z)|\le C(1+|z|)^{-\gamma},\,\,\,\,z\in\Z^d.
 \ee
%%------------------------
{\bf I2}. For every fixed $r\in\R^d$, the matrix-valued function
$R$ satisfies
 \beqn\nonumber%%\la{3.3}
 R^{ii}(r,\cdot,\cdot)\geq 0,\,\,\,\,
R^{ij}(r,x,y)=(R^{ji}(r,y,x))^T,\,\,\,\, x,y\in\Z^d_+.
\eeqn
In particular, for every fixed $r\in\R^d$,
%${\bf R}_0^{ij}(r,z)=({\bf R}_0^{ji}(r,-z))^T$,
$\hat{\bf R}_0(r,\theta)$ satisfies
\be\la{2.15'}
 \hat {\bf R}_0^{00}(r,\theta)\ge0,\,\,
\hat {\bf R}_0^{11}(r,\theta)\ge0,\,\,
\hat {\bf R}_0^{01}(r,\theta)=\hat {\bf R}_0^{10}(r,\theta)^*,
\,\,\,\,\theta\in \T^d.
\ee
{\bf I3}.  For every fixed $r\in\R^d$ and $\theta\in \T^d$,
the matrix $\hat{\bf R}_0(r,\theta)$ is nonnegative definite.\medskip\\
{\bf I4}.  For every $\theta\in \T^d$, $\hat {\bf R}_0^{ij}(\cdot,\theta)$,
$i,j=0,1$, are $C^d$ functions and the function
\be\la{2.15''}
r\to \sup_{\theta\in \T^d}\max_{i,j=0,1}
\max_{\al=(\al,\dots,\al_d):|\al_j|\le1}
\Big|\frac{\pa^\al}{\pa r_1^{\al_1}\dots\pa r_d^{\al_d}}
\hat {\bf R}_0^{ij}(r,\theta)\Big|
\ee
is bounded uniformly on bounded sets.
%In \cite{DPST} the matrices satisfying (\ref{3.2'})--(\ref{2.15''})
%are called spectral density matrix functions.
%%----------------------------------
\begin{remark}
{\rm For simplicity of proof, we could assume that
$R(r,x,y)$ has the simpler form, namely,
\be\la{Rpar}
R(r,x,y)=\zeta(x_1)\zeta(y_1){\bf R}_0(r,x-y),
\ee
where $\zeta(x)$, $x\in\Z^1$, is a nonnegative bounded function such that
$\zeta(x)=0$ for $x\le 0$ and $\zeta(x)=1$ for $x>a$ with
some $a\ge1$,  and ${\bf R}_0(r,x)$ satisfies conditions
 (\ref{3.2'})--(\ref{2.15''}). 
Then $R(r,x,y)$ satisfies {\bf I1}--{\bf I4}.
However, the limit covariance $Q_\infty(x,y)$ in (\ref{1.13})
has not the form (\ref{Rpar}), in general. 
Formula (\ref{1.13}) implies that
 $Q_\infty(x,y)$ satisfies the bound similar to (\ref{3.14}), 
$$
Q_\infty(x,y)={\bf Q}_\infty(x_1,y_1,\bar x-\bar y),\,\,
\lim_{y_1\to+\infty}{\bf Q}_\infty(y_1+z_1,y_1,\bar z)=
q_\infty(z)+q_\infty(\ti z),\,\,\,z\in\Z^d.
$$
Therefore, we will prove the main results 
under condition (\ref{3.14}) which is weaker than (\ref{Rpar}). 
}\end{remark}
%%--------------------------------------

Let $E^\ve_0$ stand for expectation with respect to the measure
$\mu_0^\ve$. Assume that
$ E^\varepsilon_0\big(Y^j(x)\big)=0$
and define the covariance
 \beqn\nonumber
 Q^{ij}_{\ve}(x,x')= E^\ve_0\big(Y^i(x)\otimes
Y^j(x')\big),\,\,\,\,x,x'\in \Z^d_+,\,\,\,\,i,j=0,1\,.
 \eeqn
%%%%%%%%%%%%%%%%%%%%%%%%%%%%%%%%%%%%%%%%
\begin{definition}
We call a family of measures $\{\mu_0^\ve,\ve>0\}$
 a family of slow variation  for $R$ if
$\{Q^{ij}_{\ve}(x,x'),\,\ve>0\}$ satisfies conditions
{\bf V1}--{\bf V3} listed below.
\smallskip\\
{\rm
% {\bf V0}.  $Q^{ij}_{\ve}(x,x')=\zeta(x_1)\zeta(x'_1){\bf Q}_\ve(x,x')$
%${\bf Q}_\ve(x,x')$ is a covariance matrix of some measure $\nu_0^\ve$
%of zero mean defined on the space ${\cal H}_\al$.
%\smallskip\\
{\bf V1}. For any $\ve >0$, there exists an even integer $N_\ve$ such that
\medskip\\
(i) for all $M\in \R^d$ and $x,x'\in I_{M}\cap\Z^d_+$,  then
\beqn\la{2.4}
\left|Q^{ij}_{\ve}(x,x')- R^{ij}(\ve M,x,x')\right|\le
C\min[(1+|x-x'|)^{-\gamma}, \ve N_\ve],
\eeqn
where $C$, $\gamma$ are the constants from (\ref{3.2}),
 and $I_{M}$ is the cube
centered at the point $M$ with edge length $N_\ve$,
 \be \la{cube}
I_{M} =\{x=(x_1,\dots,x_d)\in\Z^d:\, |x_j-M_j|\le
N_\ve/2,\,M=(M_1,\dots,M_d)\}.
 \ee
(ii) $N_\ve\sim \ve^{-\beta}$ as $\ve\to 0$, with some
$\beta\in(1/2,1)$.\smallskip\\
{\bf V2}. For any $\ve>0$ and all $x,x'\in\Z^d_+$, $i,j=0,1$, 
$
|Q^{ij}_{\ve}(x,x')|\le C(1+|x-x'|)^{-\gamma}
$
with constants $C$, $\gamma$ as in (\ref{3.2}).
\medskip

To prove the weak convergence of the measures (Theorem \ref{the3} below)
we need the stronger condition {\bf V3}:\\
{\bf V3}. The measures $\mu_0^\ve$ satisfy the Ibragimov mixing condition 
{\bf S4} (see section \ref{sec2.1})
with the mixing coefficients $\varphi_\ve$.
Moreover, it is assumed that
$$
\sup_{\ve>0}|\varphi_\ve^{1/2}(x)|\le C(1+|x|)^{-\gamma},\quad
\mbox{with a }\, \gamma>d.
$$
Note that condition {\bf V3} implies {\bf V2}.
}
\end{definition}
%%%%%%%%%%%-------------------------------------------
%For any $\ve >0$ and any $\Psi_1, \Psi_2\in{\cal S}$ with
%dist$(\supp \Psi_1,\supp \Psi_2)\ge\rho>0$ there exist constants
%$C>0$ and $\kappa\in(0,1)$  such that 
%\beqn\nonumber 
%\left| E^\ve_0 \big(e^{i\langle
%Y,\Psi_1\rangle} e^{i\langle Y,\Psi_2\rangle}\big)
%-E^\ve_0 \big(e^{i\langle Y,\Psi_1\rangle}\big)
%E^\ve_0\big(e^{i\langle Y,\Psi_2\rangle}\big) \right| \le C(1+\rho)^{-\kappa}.
%\eeqn
%For correlation functions of the fourth order
%$$M^{(4)}_\ve(x^1,x^2,x^3,x^4) = E^\ve_0\big(Y(x^1)\otimes
%Y(x^2)\otimes Y(x^3)\otimes Y(x^4)\big),\,\,\,\,x^1,\dots,x^4\in\Z^d_+,$$
%we require that
%\beqn\nonumber |M^{(4)}_\ve(x^1,x^2,x^3,x^4)|\le C
%\sum\limits_{(i_1,i_2,i_3,i_4)\in P\{1,2,3,4\}}
%(1+|x^{i_1}-x^{i_2}|)^{-\gamma}(1+|x^{i_3}-x^{i_4}|)^{-\gamma},
%\eeqn where $P\{1,2,3,4\}$ is a permutation of the numbers $1,2,3,4$,
%and $\gamma>d$.
%%%%%%%%%%%%%%%%%%%%%%%%%%%%%%%%%%%%%%%%%%%%%%%%%%%%%%%%%
\begin{definition}\la{d2.9}
(i) $\mu^\ve_t$ is a Borel probability measure on ${\cal H}_{\al,+}$
which gives the distribution of $Y(t)$,
$$
\mu^\ve_t(B) = \mu_0^\ve(U_+(-t)B),\,\,\,\mbox{where }\, 
B\in {\cal B}({\cal H}_{\al,+}) \,\,\, \mbox{and }\,t\in \R\,.
$$ 
(ii) The correlation functions of the  measure $\mu^\ve_t$ are  defined by
$$
Q_{\ve,t}^{ij}(x,y)=\int \left(Y^i(x)\otimes Y^j(y)\right)
\mu^\ve_{t}(dY)= E^{\ve}_0\big(Y^i(x,t)\otimes  Y^j(y,t)\big), 
\,\,\,i,j= 0,1,\,\,\,\,x,y\in\Z^d_+. 
$$ 
Here $Y^i(x,t)$ are  the components of the random solution 
$Y(t)=(Y^0(\cdot,t),Y^1(\cdot,t))$ to the problem (\ref{CP1}).
\end{definition}

%%%%%%%%%%%%%%%%%%%   2.3   %%%%%%%%%%%%%%%%%%%%%%
\subsection{Covariance in the kinetic scaling limit}
%%%%%%%%%%%%%%%%%%%%%%%%%%%%%%%%%%%%%%%%%%
Let us use a time span of order  $\tau/\varepsilon^\al$, $\tau\neq 0$,
with $0<\al\le1$,
and study the asymptotics of the covariance $Q_{\ve,\tau/\ve^\al}(x,y)$
as $\ve\to+0$. 
For $0<\alpha<1$, the result is given by Theorem~\ref{thea}.

%%-----------------------
%The family $\mu^\varepsilon_0$, $\varepsilon>0$, of initial
%measures has slow spatial variation on scale $\varepsilon^{-1}$
%and for long times, roughly of order $\varepsilon^{-\gamma}$,
%$0<\gamma<1$, in essence Theorem \ref{the1} applies locally, which
%implies that locally the projected measure is attained. This
%measure is then almost invariant under the time evolution. Thus
%one needs a time span of order $\tau/\varepsilon$, $\tau\neq 0$,
%to see changes in the projected part of the covariance.
%%%------------------------------

To formulate the result for $\al=1$
let us introduce the matrix $q_{\tau,r}(z)$, $z\in\Z^d_+$, 
$r\in\R^d$, $\tau\not=0$, by the  Fourier transform,
$q_{\tau,r}(z)=F^{-1}_{\theta\to z}[\hat q_{\tau,r}(\theta)]$,
where 
\beqn\la{qtaur}
\hat q_{\tau,r}(\theta)= \sum_{\si=1}^{s}
\Pi_\si(\theta)\Big[{\bf M}_+^{\sigma}(\tau;r,\theta)+
i{\bf M}_-^{\sigma}(\tau;  r,\theta)\Big]\Pi_\si(\theta),
 \quad \theta\in \T^d\setminus{\cal C}_*.
\eeqn
Here $\Pi_\si(\theta)$ is the spectral projection
introduced in Lemma \re{lc*} (iv),
\beqn\la{Pi}
\ba{lll}
 {\bf M}^\sigma_{+}(\tau;r,\theta)&=&
\frac{1}2\big({\bf R}^\sigma_+(\tau;r,\theta)+
C(\theta){\bf R}^\sigma_+(\tau;r,\theta)C^*(\theta)\big),\medskip\\
{\bf M}_{-}^\sigma(\tau;r,\theta)&=&\frac{1}2\big(C(\theta)
{\bf R}_{-}^\sigma(\tau;r,\theta)
-{\bf R}_{-}^{\sigma}(\tau;r,\theta)C^*(\theta)\big),
\ea
\eeqn
with $C(\theta)$ defined in (\ref{C(theta)}),
and 
\beqn\la{gclimcor2}
{\bf R}^\sigma_\pm(\tau;r,\theta)=\frac{1}2
\Big(\hat {\bf R}_0(r+\nabla\om_\sigma(\theta)\tau,\theta)
\chi^+_{\tau,r_1}(\theta)
\pm\hat {\bf R}_0(r-\nabla\om_\sigma(\theta)\tau,\theta)
\chi^-_{\tau,r_1}(\theta)\Big),
\eeqn
where 
\beqn\la{chi}
\chi^\pm_{\tau,r_1}(\theta)=
\frac12(1+\sign(r_1\pm\tau\nabla_1\om_\sigma(\theta)).
\eeqn
%%%%%%%%%%%%%%%%%%%%%%%%%%%%%%%%
\begin{theorem}\label{the2'}
 Let conditions {\bf V1}--{\bf V2} and {\bf E1}--{\bf E6} hold.
Then for any $\tau\not=0$, $r\in\R^d$ with $r_1\ge0$, 
the correlation functions converge to a limit,
\beqn\la{gclimcor0}
 \lim_{\ve\to 0}Q_{\ve,\tau/\ve}([r/\ve]+z,[r/\ve]+z')
=Q^G_{\tau,r}(z,z'),
\eeqn
where 
$$
Q^G_{\tau,r}(z,z')=\left\{
\ba{lr}
q^G_{\tau,r}(z-z')=q_{\tau,r}(z-z')+q_{\tau,\tilde r}(\ti z-\ti z'),
& \mbox{if }r_1>0,\,\,z,z'\in\Z^d;\\
q_{\tau,r}(z\!-\!z')-q_{\tau,r}(z\!-\!\tilde z')-
q_{\tau,r}(\tilde z\!-\!z')+q_{\tau,r}(\tilde z\!-\!\tilde z'),
& \mbox{if }r_1=0, \,\,z,z'\in\Z^d_+,\ea\right.
$$
with $q_{\tau,r}(z)$ defined by (\ref{qtaur})--(\ref{gclimcor2}).
\end{theorem}
%%-------------------------------------------

This theorem is proved in Section \ref{sec.3}.
%%%---------------------------------------
\begin{cor}
Let  $r_1>0$ and $\tau\not=0$.
(i) From formulas (\ref{qtaur})--(\ref{gclimcor2}) it follows that 
the $\sigma$-band of $\hat q^G_{\tau,r}(\theta)$ 
satisfies the following "hydrodynamic" equation:
$$
\partial_\tau f(\tau,r;\theta)=iC_\sigma(\theta)
\nabla\omega_\sigma(\theta)
\cdot \nabla_r f(\tau,r;\theta),\,\,\,\,r_1>0,\,\,\,\tau>0,
$$
where $C_\sigma=\left(\ba{cc}0&\omega^{-1}_\sigma\\
-\omega_\sigma&0\ea\right),\,\,\, \sigma=1,\dots,s,
$
 the boundary and initial conditions are given by the $\hat {\bf R}_0$.\\
%%%%-------------------------
(ii) From formulas (\ref{qtaur})--(\ref{gclimcor2})
it follows that $q^G_{\tau,r}$ satisfies the equilibrium condition
(\ref{2.10'}), i.e., $\hat q^{G,11}_{\tau,r}(\theta)^*=
\Omega^2(\theta)\hat q^{G,00}_{\tau,r}(\theta)$,
$\hat q^{G,01}_{\tau,r}(\theta)=-\hat q^{G,10}_{\tau,r}(\theta)$.
Moreover, 
$%\hat q^{G,ii}_{\tau,r}(\theta)^*=
\hat q^{G,ii}_{\tau,r}(\theta)\ge0$,
$\hat q^{G,01}_{\tau,r}(\theta)^*=\hat q^{G,10}_{\tau,r}(\theta)$.
\end{cor}
%%-----------------------------

Let us introduce the scaled $n\times n$ Wigner
matrix through
 \beqn\nonumber%\la{3.15}
W^\varepsilon(\tau;r,\theta)=
\sum_{y\in \Z^d}e^{i\theta\cdot y}\,
E^\varepsilon_{\tau/\varepsilon}
\Big(a^\ast([r/\varepsilon+y/2])
\otimes a([r/\varepsilon-y/2])\Big),\quad r\in \R^d_+,
 \eeqn
where $a(x)$ is given in (\ref{2.1}).
By conditions {\bf V1} and {\bf V2}, the following limit exists
 \beqn
 \lim_{\varepsilon\to 0}W^\varepsilon(0;r,\theta)&=&
 \frac{1}{2}\Big(\Omega^{1/2}\hat {\bf R}_0^{00}(r,\theta)\Omega^{1/2}+
 \Omega^{-1/2}\hat {\bf R}_0^{11}(r,\theta)\Omega^{-1/2}\nonumber\\
 &&+i\Omega^{1/2}\hat {\bf R}_0^{01}(r,\theta)\Omega^{-1/2}-i
 \Omega^{-1/2}\hat {\bf R}_0^{10}(r,\theta)\Omega^{1/2}\Big)\nonumber\\
 &=& W(0;r,\theta),\quad\, r_1>0.\nonumber
 \eeqn
%%%%$W^\varepsilon(0;r,\theta)=0$ if $r_1=0$.

We also define the limit  Wigner matrix as follows.
 \beqn\la{3.16a}
 W^{p}(\tau;r,\theta)\!\!&=&\!\!\sum^s_{\sigma=1}
 \Pi_\sigma(\theta)
\Big\{W(0;r-\tau\nabla\omega_\sigma(\theta),\theta)
\chi^-_{\tau,r_1}(\theta)\nonumber\\
&&\!\!\!\!+W(0;-r_1+\tau\nabla_1\omega_\sigma(\theta),
\bar r-\tau\bar\nabla\omega_\sigma(\theta),\ti\theta)
\chi^+_{\tau,-r_1}(\theta)\Big\}\Pi_\sigma(\theta)\\
\!\!&=&\!\!\left\{ \ba{ll}
\sum\limits^s_{\sigma=1} \Pi_\sigma(\theta)
W(0;r\!-\!\tau\nabla\omega_\sigma(\theta),\theta)
\Pi_\sigma(\theta),\!\!&\!\mbox{if }\,r_1>\tau\nabla_1\omega_\sigma(\theta),\\
\sum\limits^s_{\sigma=1} \Pi_\sigma(\theta)
W(0;-\!r_1\!+\!\tau\nabla_{\theta_1}\omega_\sigma(\theta),
\bar r\!-\!\tau\nabla_{\bar \theta}\omega_\sigma(\theta),\tilde\theta)
\Pi_\sigma(\theta),\!\!&\!\mbox{if }\,r_1<\tau\nabla_1\omega_\sigma(\theta),
\ea
\right.\nonumber
 \eeqn
where $\ti\theta=(-\theta_1,\bar\theta)$, 
$\bar \theta=(\theta_2,\dots,\theta_d)$,$\bar r=(r_2,\dots, r_d)$.
%%%%%%%%%%%%%%%%%%%%%%%%%%%%%%%%%%%%
\begin{theorem} \la{the2}
Let conditions {\bf{V1}}--{\bf{V2}} 
and {\bf{E1}}--{\bf{E6}} hold. Then for any $r\in\R^d_+$ and
$\tau>0$, the following limit exists in the sense of distributions,
 \be\la{3.17}
 \lim_{\varepsilon\to 0} W^\varepsilon(\tau;r,\theta)=
 W^{p}(\tau;r,\theta)\,.
 \ee
In addition, for the remaining part of the covariance,
$$
 \lim_{\varepsilon\to 0}\sum_{y\in\Z^d}
e^{i\theta\cdot y}\,E^\varepsilon_{\tau/\varepsilon}
\Big(a([r/\varepsilon+y/2])\otimes  a([r/\varepsilon-y/2])\Big)=0\,.
 $$
\end{theorem}
%%%%%%%%%%%%%%%%%%%%%%%%%%%%%%%%%%%%%%%%%%%%%%

This theorem  is proved in Section \ref{sec.4}.

%We remark that in the $\sigma$-th band of the Wigner function evolves
%according to the transport equation
% \be\la{3.18}
% \frac{\partial}{\partial  t}f_t(r,\theta)
%+\nabla\omega_\sigma(\theta)\cdot \nabla_r
% f_t(r,\theta)=0\,,\quad r\in\R^d_+,
% \ee
%where the boundary and initial conditions are given by the initial Wigner
%matrix projected onto the $\sigma$-th band.
%%%%%%%%--------------------------------------------------
\begin{cor}\la{cor2.13}
Denote by 
$W_{\sigma}^{p}(\tau;r,\theta)$,
$\sigma=1,\dots,s$, the $\sigma$-th band of the Wigner function
$W^{p}(\tau;r,\theta)$.
Then $W_{\sigma}^{p}$ is a solution of the "energy transport" equation 
$$
\pa_\tau W_{\sigma}^{p}(\tau;r,\theta)
+\nabla\om_\sigma(\theta)\cdot\nabla_rW_{\sigma}^{p}(\tau;r,\theta)
=0,\,\,\,\tau>0,\,\,\,r\in\R^d_+,
$$
where the boundary and initial conditions are given by the initial Wigner
matrix projected onto the $\sigma$-th band,
\beqn
W_{\sigma}^{p}(\tau;r,\theta)|_{\tau=0}&=&
\Pi_\sigma(\theta) W(0;r,\theta)\Pi_\sigma(\theta),\,\,\,r\in\R^d_+,
\nonumber\\
 W_{\sigma}^{p}(\tau;r,\theta)|_{r_1=0}&=&b(\tau;\bar r,\theta),
\quad \bar r\in\R^{d-1}, \quad \tau>0.\nonumber
\eeqn
Here 
$$
b(\tau;\bar r,\theta):=
\left\{\ba{cc}
\Pi_\sigma(\theta)W(0;(-\tau\nabla_1\omega_\sigma(\theta),
\bar r-\tau\bar\nabla\omega_\sigma(\theta)),\theta)\Pi_\sigma(\theta),&
\mbox{if }\,\nabla_1\om_\sigma(\theta)<0,\\
\Pi_\sigma(\theta)W(0;(\tau\nabla_1\omega_\sigma(\theta),
\bar r-\tau\bar\nabla\omega_\sigma(\theta)),\ti\theta)\Pi_\sigma(\theta),&
\mbox{if }\,\nabla_1\om_\sigma(\theta)>0.
\ea\right.
$$
\end{cor}

%%%%%%%%%%%%%%%%%%%  2.4   %%%%%%%%%%%%%%%%%%%
\subsection{Weak convergence of measures family}
%%%%%%%%%%%%%%%%%%%%%%%%%%%%%%%%%%%%
Let us consider the random field $Y$ at the
kinetic time $\tau/\varepsilon$, $\tau\neq 0$, and close to the
spatial point $[r/\varepsilon]\in \Z^d_+$. Denote by
$T_h$, $h\in \Z^d_+$, the group of space
translations. The measure at $r/\varepsilon$ is then defined
through
 \be\la{13.19}
 \mu^\varepsilon_{\tau/\varepsilon,r}=T_{-[r/\varepsilon]}
 \mu^\varepsilon_{\tau/\varepsilon}\,,
 \ee
i.e., $\mu^\varepsilon_{\tau/\varepsilon,r}(B)=
 \mu^\varepsilon_{\tau/\varepsilon}(T_{[r/\varepsilon]}B)$,
where $B\in{\cal B}({\cal H}_{\al,+})$
and $\mu_{\tau/\varepsilon}^\ve$ is defined in Definition \ref{d2.9}.
%%%%%%%%%%%%%%%%%%%%%%%%%%%%%%%%%%%%%%%%%%%
\begin{theorem}\label{the3}
 Let conditions {\bf{V1}}--{\bf{V3}} and {\bf{E1}}--{\bf{E6}} hold. 
Then for $\tau\neq 0$, $r\in\R^d$ with $r_1\ge0$,
in the sense of weak convergence on ${\cal H}_{\alpha,+}$,
 \be\la{13.20}
 \lim_{\varepsilon\to 0}\mu^\varepsilon_{\tau/\varepsilon,r}
=\mu^{G}_{\tau,r}\,.
 \ee
The measure  $\mu^{G}_{\tau,r}$ is a Gaussian measure 
on ${\cal H}_{\alpha,+}$, which is invariant under the time
translation $U_+(t)$. $\mu^{G}_{\tau,r}$ has mean zero and covariance
 \beqn\nonumber
Q^{G,ij}_{\tau,r}(z,z')=
 \int\big(Y^i(z)\otimes Y^j(z')\big)\,\mu^{G}_{\tau,r}(dY),
 \eeqn
defined by Theorem \ref{the2'}.
If $r_1>0$, the covariance 
$Q^{G,ij}_{\tau,r}(z,z')=q^{G,ij}_{\tau,r}(z-z')$
is determined through $W^p(\tau;r,\theta)$ as
\be\la{13.21}
  \Omega(\theta)\hat{q}^{G,00}_{\tau,r}(\theta)=
  \Omega(\theta)^{-1}\hat{q}^{G,11}_{\tau,r}(\theta)=
  \frac{1}{2}\big(W^p(\tau;r,\theta)+
  W^p(\tau;r,-\theta)^T\big)
\ee
  and
\be\la{13.22}
\hat{q}^{G,01}_{\tau,r}(\theta)= -\hat{q}^{G,10}_{\tau,r}(\theta)=
  -\frac{i}{2}\big(W^p(\tau;r,\theta)- W^p(\tau;r,-\theta)^T\big)\,.
\ee
\end{theorem}
%%--------------------------------------------------------

Theorem \ref{the3}  is proved  in Section \ref{sec.5}.

From Theorem \ref{the3} 
we conclude that close to $r/\varepsilon$ in space and close to
$\tau/\varepsilon$ in time the random field $Y^j(x,t)$ is a
stationary Gaussian field. Its distribution at fixed local time
$\tau$ is given by $\mu^G_{\tau,r}$ while in time it evolves
deterministically according $U_+(t)$. In this sense locally in space
and time the random field is stationary with statistics determined
through the Wigner matrix at $(r,\tau)$ and the microscopic
dynamics, compare with (\ref{13.21}), (\ref{13.22}).
%%-----------------------------------
\medskip

Let us use a time span of order $\tau/\ve^{\alpha}$
with  an $\alpha\in(0,1)$. 
In this case,  change condition~{\bf V1} (ii) as follows:
$N_\ve\sim\ve^{-\beta}$ as  $\ve\to0$ with some $\beta\in(\al/2,\al)$.
Then the following result holds.
%%------------------------------------------------------
\begin{theorem}\la{thea}
 Let $\alpha\in(0,1)$, conditions {\bf V1}--{\bf V2} 
and {\bf E1}--{\bf E6} hold.
Then for $\tau\not=0$, \\
(i) the correlation functions of
measures $\mu^\ve_{\tau/\ve^\al,r}$ converge to a limit,
$$
 \lim_{\ve\to 0}Q_{\ve,\tau/\ve^\al}([r/\ve]+z,[r/\ve]+z')
=Q_{r}(z,z'),\quad z,z'\in\Z^d,
$$
where $Q_{r}(z,z')$ does not depend on $\tau$ and has the form
$$
Q_{r}(z,z')=\left\{\ba{lr}
q_{r}(z-z'),& \mbox{if }r_1>0;\\
q_{r}(z-z')-q_{r}(z-\tilde z')-
q_{r}(\tilde z-z')+q_{r}(\tilde z-\tilde z'),
& \mbox{if }r_1=0;\\
\ea\right.
$$
where in Fourier space, 
$
\hat q_{r}(\theta)=
\sum_{\si=1}^{s}\Pi_\si(\theta) M_{r}(\theta)\Pi_\si(\theta),
$
with
\beqn\nonumber
%%M_{r}(\theta)&=& 0,\,\mbox{if }\,r_1<0;\nonumber\\
M_{r}(\theta)&=&\frac12\Big[
\hat {\bf R}_0(r,\theta)
+C(\theta)\hat {\bf R}_0(r,\theta)C^*(\theta)\Big], \,\mbox{if } {r_1>0};
\nonumber\\
M_{r}(\theta)&=&\frac14\Big[\hat {\bf R}_0(r,\theta)
+C(\theta)\hat {\bf R}_0(r,\theta)C^*(\theta)\Big]
\nonumber\\
&&+i\Big(C(\theta)\hat {\bf R}_0(r,\theta)
-\hat {\bf R}_0(r,\theta)C^*(\theta)\Big)
\sign(\nabla_1\omega_\sigma(\theta))\Big],\,\mbox{if }\,r_1=0
\nonumber.
\eeqn
(ii) The measures $\mu^\varepsilon_{\tau/\varepsilon^{\alpha},r}$
% ( $\tau\neq 0$,$r\in\R^d_+$) 
converge weakly on the space ${\cal H}_{\alpha,+}$
to a limit measure $\mu_{r}$ as $\varepsilon\to0$. 
Moreover,  $\mu_{r}$ is a Gaussian measure on ${\cal H}_{\alpha,+}$,
which is invariant under the time
translation $U_+(t)$, has mean zero and
covariance $Q_{r}(z,z')$ defined above.
\end{theorem}
%%----------------------------------
%Note that in the case when $r_1=0$ 
%the form $\hat q_r(\theta)$ coincides with 
%$\hat q_\infty(\theta)$
%(if in formulas (\ref{1.14})--(\ref{1.15})) we set $\hat {\bf R}_0(r,\theta)$
%instead of $\hat {\bf q}_0(\theta)$);
%if $r_1>0$ $\hat q_r(\theta)$ coincides with the limit correlation functions
%for harmonic crystals in the whole space $\Z^d$
%(see \cite{DS}).
%%%%%%%%%%%%%%%%%%%%%%%%%%%%%%%%

We omit the proof of Theorem \ref{thea} since it can be proved by using
the technique of Theorems~\ref{the2'} and \ref{the3}.

%%%%%%%%%%%%%%%%%%%%%%%%%%%    3     %%%%%%%%%%%%%%%%%%%%%%%%%%%%%%%%%%
\setcounter{equation}{0}
\section{Convergence of correlation functions}\la{sec.3}
%%--------------------------------------------------
In this section we prove Theorem \ref{the2'}.
Before we outline the strategy of the proof.
At first, we use the cutting strategy from \ci{DKS1}-\cite{DS}
combined with some techniques from \ci{DPST},
where Theorem \ref{the2'} has proved for the case when
$d=n=1$ and in the entire space (see \ci[Theorem 3.1]{DPST}).
Note that in \cite{DPST} it is assumed the stronger
conditions on matrix $V$ than {\bf E3}, {\bf E4}, namely, 
 $\omega(\theta)>0$, and  
%(see condition (iv) in \cite[p.581]{DPST}) 
the set
$$
\{\theta\in[-\pi,\pi]:\,\omega''(\theta)=\omega'''(\theta)=0\}
$$
is empty.
%Instead of these conditions we assume weaker conditions:
%for $d=n=1$ $\omega(\theta)\ge0$ ({\bf E3}) and $\omega''(\theta)$
%does not vanish identically ({\bf E4}).
Under these conditions, in \ci{DPST} 
the uniform asymptotics
of the Green function is proved,
\be\la{asGrf}
\sup_{x\in\Z^d}|{\cal G}_t(x)|\le C(1+ |t|)^{-1/3}.
\ee
This bound plays an important role in the proof of \ci{DPST}.
However,
if $n > 1$, then $\omega_s$ may be non-smooth because of band crossing,
and if $d > 1$, the set where the Hessian vanishes does not consist of
isolated points.
Therefore a strong estimate as (\ref{asGrf}) is unlikely to be valid, in
general.
  To cope with such a situation, we split ${\cal G}_t(x)$
  into two summands:  ${\cal G}_t(x)={\cal G}^f_t(x)+
{\cal G}^g_t(x)$, where  ${\cal G}^f_t(x)$
has a support in the neighborhood of a  ``critical set''
${\cal C}\subset \T^d$, and ${\cal G}^g_t(x)$
vanishes in the neighborhood of ${\cal C}$.
The set ${\cal C}$ includes  all points $\theta\in \T^d$
either with a degenerate Hessian  of $\om_\sigma(\theta)$,
 or with non-smooth  $\om_\sigma(\theta)$
(see formula (\ref{calC})). We show that
 the contribution of ${\cal G}^f_t(x)$ is negligible
uniformly in $t$ (see (\ref{3.12})).
Hence, it allows us to represent correlations functions
$Q_{\ve,\tau/\ve}$ in the form:
$Q_{\ve,\tau/\ve}=Q^g_{\ve,\tau/\ve}+Q^r_{\ve,\tau/\ve}$,
such that
\beqn\nonumber
Q^g_{\ve,\tau/\ve}(x,y)=\sum\limits_{x',y'\in\Z^d}
{\cal G}^g_{t,+}(x,x')
Q_{\ve}(x',y'){\cal G}^g_{t,+}(y,y')^*.
\eeqn
For the remainder $Q^r_{\ve,\tau/\ve} =Q_{\ve,\tau/\ve}-Q^g_{\ve,\tau/\ve}$
we prove that $Q^r_{\ve,\tau/\ve}(x,y)=o(1)$
uniformly in $\tau\not=0$, $\ve>0$ and $x,y\in\Z^d$.
The last fact follows from two key observations:
(i) mes${\cal C}=0$ (Lemma \ref{lc*}) and 
(ii) the correlation quadratic form is continuous in $\ell^2_+$,
see Corollary \ref{c4.10}.
Up to this point we apply the ``cutting strategy" from \ci{DKS1}--\cite{DS}.
Finally,  we prove that
$Q^g_{\ve,\tau/\ve}([r/\ve]+x,[r/\ve]+y)$
converges to a limit as $\ve\to0$, using the techniques of \ci{DPST}
and \cite{DKM}.
In addition, the asymptotics of ${\cal G}^g_t(x)$,
(see Lemma \ref{l5.3}) of the form
${\cal G}^g_t(x)\sim (1+|t|)^{-d/2}$ plays the important role,
since it replaces the asymptotics (\ref{asGrf})
and also simplifies some steps of the proof of \ci{DPST}.
However, in our case the structure of $R(r,x,y)$ is more complex than
in \cite{DPST} or \cite{DS}, in which
$R(r,x,y)={\bf R}_0(r,x-y)$. We apply the approach of \cite{DKM, D08},
where convergence to equilibrium was proved for
 non translation-invariant initial measures,
and combine  with the technique of \cite{DS}.

%%------------------------  3.1  -----------------------------
\subsection{Bounds for initial covariance}
%%%---------------------------------------------------
\begin{definition}
By $\ell^p\equiv \ell^p(\Z^d)\otimes \R^n$ 
(by $\ell^p_+\equiv \ell^p(\Z^d_+)\otimes \R^n)$, where $p\ge 1$ and
 $n\ge 1$, denote the space of sequences
$f(z)=(f_1(z),\dots,f_n(z))$ endowed with norm
$\Vert f\Vert_{\ell^p}=\Big(\sum\limits_{z\in\Z^d}|f(z)|^p\Big)^{1/p}$,
respectively,
$\Vert f\Vert_{\ell^p_+}:=\Big(\sum\limits_{z\in\Z^d_+}|f(z)|^p\Big)^{1/p}$.
\end{definition}

The following lemma follows from condition {\bf V2}.
%%%%%%%%%%%%%%%%%%%%%%%%%%%%%%%%%%%%%%%%%%%%%%%%%%%%%%%%%%%%%%%%%
\begin{lemma} \la{l4.1}
Let  condition {\bf V2} hold. Then, 
for $i,j=0,1$, the following bounds hold:
\beqn
\sum\limits_{z'\in\Z^d_+} |Q^{ij}_\ve(z,z')|
&\le& C<\infty\,\,\,\mbox{ for all }\,z\in\Z^d_+,%\la{pr1}
\nonumber\\
\sum\limits_{z\in\Z^d_+} |Q^{ij}_\ve(z,z')|
&\le& C<\infty\,\,\,\mbox{ for all }\,z'\in\Z^d_+.\la{pr2}
\nonumber
\eeqn
Here the constant $C$ does not depend on $z,z'\in \Z^d_+$
and $\ve>0$.
\end{lemma}
%%%%%%%%%%%%%%%%%%%%%%%%%%%%%%%%%%%%%%%
\begin{cor}\la{c4.10}
By the Shur lemma, it follows from Lemma \ref{l4.1} that  
\beqn\nonumber
|\langle Q_\ve(z,z'),\Phi(z)\otimes\Psi(z')\rangle_+|\le
C\Vert\Phi\Vert_{\ell^2_+} \Vert\Psi\Vert_{\ell^2_+},\,\,\,\,
\mbox{for any }\,\,\Phi,\Psi\in \ell^2_+,
\eeqn
where the constant $C$ does not depend on  $\ve>0$.
\end{cor}

%%%%%%%%%%%%%%%%%%%%%%%%%%%%%%%%%  3.2  %%%%%%%%%%%%%%%%%%%%%
\subsection{Stationary phase method}
%%%%%%%%%%%%%%%%%%%%%%%%%%%%%%%%%%%%%%%%%%%%%%%%%%

By (\ref{Grcs}) and (\ref{hA}) we see that $\hat{\cal G}_t( \theta)$ is
of the form 
\beqn\la{4.5}
\hat{\cal G}_t( \theta)=
\left( \begin{array}{cc}
 \cos\Om t &~ \sin \Om t~\Om^{-1}  \\
 -\sin\Om t~\Om
&  \cos\Om t\end{array}\right),
\eeqn
where $\Om=\Om( \theta)$ is the
Hermitian matrix defined by (\ref{Omega}).
Hence, we can rewrite ${\cal G}_t(x)$ in the form
\be\la{3.4'}
{\cal G}_t(x)=
\sum\limits_{\pm,\sigma=1}^s \int\limits_{\T^d}
e^{-ix\cdot \theta}e^{\pm i\om_\sigma(\theta)\,t}
a^\pm_\sigma(\theta)\,d\theta,
\ee
by (\ref{spd'}).
We are going to apply the stationary phase arguments
to the integral (\ref{3.4'}) which require a smoothness
in $\theta$. Then we have to choose certain smooth
branches of the functions $a^\pm_\sigma(\theta)$
 and $\om_\sigma(\theta)$ and cut off all singularities.
First, introduce the {\it critical set} as
\beqn\la{calC}
{\cal C} ={\cal C}_0\bigcup
{\cal C}_*\bigcup_{\sigma=1}^{s}\left({\cal C}_\sigma
 \bigcup\limits_{i=1}^{d}\,
\Big\{\theta\in \T^d\setminus {\cal C}_*:\,
 \frac{\pa^2\omega_\sigma(\theta)}{\pa \theta_i^2}=0\Big \}
\bigcup\{\theta\in \T^d\setminus {\cal C}_*:\,
\nabla_1\omega_\sigma(\theta)=0\}\right),
\eeqn
with ${\cal C}_*$ as in Lemma \ref{lc*} and sets ${\cal C}_0$
and ${\cal C}_\sigma$ defined by (\ref{c0ck}).
Obviously, ${\rm mes}\,{\cal C}=0$ (see \cite[lemma 7.3]{DKM}).
  Secondly, fix an $\delta>0$ and choose a finite partition of unity,
\beqn\la{part}
f(\theta)+ g(\theta)=1,\,\,\,\,
g(\theta)=\sum_{k=1}^K g_k(\theta),
\,\,\,\,\theta\in \T^d,
\eeqn
where $f,g_k$ are nonnegative functions in $C_0^\infty(\T^d)$, and
%%%%%%%%%%%%%%%%%%%%%%%%%%%%%%%%%%%%%%%%%%%%%%%%%%
\be\la{fge}
\supp f\subset \{\theta\in \T^d:\,
{\rm dist}(\theta,{\cal C})<\delta\},\,\,\,
\supp g_k\subset \{\theta\in \T^d:\,
{\rm dist}(\theta,{\cal C})\ge\delta/2\}.
\ee
Then we represent ${\cal G}_t(x)$ in the form
%%\la{3.10}
${\cal G}_t(x)={\cal G}^f_t(x)+{\cal G}^g_t(x)$,
where
\beqn\la{frepecut}
{\cal G}^f_t(x)&=&(2\pi)^{-d}\int\limits_{\T^d}
e^{-ix\cdot\theta}f(\theta)\,\hat{\cal G}_t(\theta)\,d\theta,\\
\la{frepecut'}
{\cal G}^g_t(x)&=&(2\pi)^{-d}\int\limits_{\T^d}
e^{-ix\cdot\theta}g(\theta)\,\hat{\cal G}_t(\theta)\,d\theta
=\sum\limits_{\pm,\sigma=1}^s\sum\limits_{k=1}^{K} \int\limits_{\T^d}
g_k(\theta)e^{-ix\cdot\theta\pm i\om_\sigma(\theta)t}
a^\pm_\sigma(\theta)\,d\theta.
\eeqn
By Lemma \re{lc*} and the compactness arguments,
we can choose the supports of $g_k$ so small that the eigenvalues 
$\om_\sigma(\theta)$ and the amplitudes $a^\pm_\sigma(\theta)$
are real-analytic functions inside
the $\supp g_k$ for every $k$. (We do not label the
functions by the index $k$ to simplify the notation.)
For the function ${\cal G}^f_t(x)$,
the Parseval identity, (\ref{4.5}), and condition {\bf E6} imply
\be\la{3.12}
\Vert {\cal G}^f_t(\cdot)\Vert^2_{\ell^2}=C
\int\limits_{\T^d} |\hat{\cal G}_t(\theta)|^2|f(\theta)|^2\,d\theta
\le C\int\limits_{{\rm dist}(\theta,{\cal C})<\delta}
|\hat{\cal G}_t(\theta)|^2\,d\theta \to 0\,\,\,\mbox{as } \,\,\de\to 0,
\ee
uniformly in $t\in\R$.
For the function ${\cal G}^g_t(x)$ the following lemma holds.
%%%%%%%%%%%%%%%%%%%%%%%%%%%%%%%%%%%%%%%%%%%%%
\begin{lemma}\la{l5.3} (see \cite[Lemma 4.5]{DS})
Let conditions {\bf E1}--{\bf  E4} and {\bf E6} hold. Then
the following bounds hold.

(i) $\sup_{x\in\Z^d}|{\cal G}^g_t(x)| \le  C~t^{-d/2}$.

(ii) For  any $p>0$, there exist $C_p,\gamma_g>0$ such that
$|{\cal G}^g_t(x)|\le C_p(|t|+|x|+1)^{-p}$ for
$|x|\ge \gamma_g t$.
\end{lemma}

%%%%%%%%%%%%%%%%%%%%%%%%%%%%%%%%%%%%%%%%%%%%%%%%%%%%
%%%%%%%%%%%%%%%%%%%%%    3.3     %%%%%%%%%%%%
\subsection{Proof of Theorem 2.10
%\ref{the2'}
}
%%%---------------------------------
 The representation (\re{sol}) yields
\beqn\la{3.4}
Q_{\ve,t}(z,z')=E^\ve_0\big(Y(z,t)\otimes Y(z',t)\big)
=\sum\limits_{x,y\in \Z^d_+}
{\cal G}_{t,+}(z,x)Q_\ve(x,y){\cal G}_{t,+}(z',y)^T,\,\,\, z,z'\in\Z^d_+,
\eeqn
for any $t\in\R^1$.
It follows from condition (\ref{condE0}) and from formulas
(\ref{Grcs}) and (\ref{hA}) that ${\cal G}_t(z)={\cal G}_t(\ti z)$
with $\ti z=(-z_1,z_2,\dots,z_d)$. In this case, by (\ref{sol1}),
the covariance $Q_{\ve,t}(z,z')$ can be decomposed into the
 sum of fourth terms,
$$
Q_{\ve,t}(z,z')=S_{\ve,t}(z,z')-S_{\ve,t}(\ti z,z')-S_{\ve,t}(z,\ti z')
+S_{\ve,t}(\ti z,\ti z'),\quad z,z'\in\Z^d_+,
$$
where
\beqn
S_{\ve,t}(z,z'):=\sum\limits_{x,y\in \Z^d_+}
{\cal G}_{t}(z-x)Q_\ve(x,y){\cal G}_{t}(z'-y)^T.
\eeqn
%%----------------------------------------------
\begin{pro}\la{p4}
Let $r\in\R^d$ and $z,z'\in\Z^d$. Then
\be\la{pro4}
S_{\ve,\tau/\ve}([r/\ve]+z,[r/\ve]+z')\to q_{\tau,r}(z-z'),\quad \ve\to+0,
\ee
where $q_{\tau,r}(z)$  is defined in 
(\ref{qtaur})--(\ref{gclimcor2}).
\end{pro}
%%---------------------------------------

This proposition implies Theorem \ref{the2'}. Indeed, 
let $r_1=0$. Then $\ti r/\ve=r/\ve=(0,\bar r/\ve)$ and
for $z,z'\in\Z^d_+$,
\beqn
Q_{\ve,\tau/\ve}([r/\ve]+z,[r/\ve]+z')&=&
S_{\ve,\tau/\ve}([r/\ve]+z,[r/\ve]+z')
-S_{\ve,\tau/\ve}([r/\ve]+\ti z,[r/\ve]+z')
\nonumber\\
&&-S_{\ve,\tau/\ve}([r/\ve]+z,[r/\ve]+\ti z')
+S_{\ve,\tau/\ve}([r/\ve]+\ti z,[r/\ve]+\ti z').\nonumber
\eeqn
Therefore, convergence (\ref{pro4}) implies (\ref{gclimcor0}).

Let $r_1>0$. In this case,
the matrix-valued functions 
$S_{\ve,\tau/\ve}([\ti r/\ve]+\ti z,[r/\ve]+z')$ and 
$S_{\ve,\tau/\ve}([r/\ve]+z,[\ti r/\ve]+\ti z')$ 
vanish as $\ve\to+0$, and 
$$
S_{\ve,\tau/\ve}([\ti r/\ve]+\ti z,[\ti r/\ve]+\ti z')
\to q_{\tau,\ti r}(\ti z-\ti z'),\quad \ve\to+0.
$$
It can be proved by similar way as Proposition \ref{p4}.
\medskip\\
%%%%----------------------------------------------------
{\bf Proof of Proposion \ref{p4}}.
{\it Step (i)} Let us denote
$$
\bar Q_\ve(x,y)=\left\{
\ba{cl} Q_\ve(x,y)&\mbox{for }\, x,y\in\Z^d_+,\\
0& \mbox{otherwise}
\ea\right.
$$
Corollary \ref{c4.10} and (\ref{3.12}) imply that
\beqn\nonumber
S_{\ve,t}(z,z')
=\sum\limits_{x,y\in \Z^d}{\cal G}^g_t(z-x)\bar Q_\ve(x,y)
{\cal G}^{g}_t(z'-y)^T+o(1),
\eeqn
where $o(1)\to 0$ as  $\delta\to 0$ uniformly in $t\in \R$ and $z,z'\in\Z^d$.
In particular, setting $t=\tau/\ve$,
$z=[r/\ve]+l$ and $z'=[r/\ve]+p$ we get
\beqn
S_{\ve,\tau/\ve}([r/\ve]\!+\!l,[r/\ve]+p)&=&
\sum\limits_{x,y\in \Z^d}
{\cal G}^g_{\tau/\ve}([r/\ve]+l-x)
\bar Q_\ve(x,y){\cal G}^{g}_{\tau/\ve}(p+[r/\ve]-y)^T+o(1)\nonumber\\
&=&\sum\limits_{x,y\in \Z^d}
{\cal G}^g_{\tau/\ve}(l+x)
\bar Q_\ve([r/\ve]\!-\!x,[r/\ve]\!-\!y){\cal G}^{g}_{\tau/\ve}(p+y)^T
+o(1).\nonumber
\eeqn
Let $c =\gamma_g+\max(|l|,|p|)$. Then  Lemma \ref{l5.3} (ii) and
condition {\bf V2}  imply that
\beqn
S_{\ve,\tau/\ve}([r/\ve]\!+\!l,[r/\ve]+p)&=&
\sum\limits_{x,y\in [-c\tau/\ve,c \tau/\ve]^d}
{\cal G}^g_{\tau/\ve}(l+\!x)
\bar Q_\ve([r/\ve]\!-\!x,[r/\ve]\!-\!y){\cal G}^{g}_{\tau/\ve}(p+y)^T
\nonumber\\
&&+r_1(\ve,\tau)+o(1),\quad \ve\to0,\nonumber
\eeqn
where $\lim\limits_{\ve\to0}\ve^{-p}r_1(\ve,\tau)=0$
for any $p>0$ and $\tau\in\R^1$.
%%%%%%%%%%%%%%%%%%%%%%%%%%%%%%%%%%%%%%

{\it Step (ii)}
We divide the cube $[-c \tau/\ve,c \tau/\ve]^d$
onto the cubes $I_{n N_\ve}$ (see (\ref{cube})),
$$
[-c \tau/\ve,c \tau/\ve]^d
 \subset\bigcup\limits_{n\in J} I_{n N_\ve},
 $$
where $J =\{n=(n_1,\dots,n_d)\in\Z^d,\,
|n_j|\le[c \tau/(\ve N_\ve)]+1\}$. Therefore,
$$
S_{\ve,\tau/\ve}([r/\ve]\!+\!l,[r/\ve]+p)=
\sum\limits_{m,n\in J}\!\!\!
\sum\limits_{\scriptsize\ba{c}
x\in I_{m N_\ve}\\
y\in I_{n N_\ve}\ea}\!\!\!\!\!\!\!
{\cal G}^g_{\tau/\ve}(l+\!x)\bar Q_\ve([r/\ve]\!-\!x,[r/\ve]\!-\!y)
{\cal G}^{g}_{\tau/\ve}(p+y)^T+o(1).
$$
Now we prove that 
the contribution of the sums over pairs $m,n\in J$ with $m\ne n$
vanishes as $\ve\to0$.
%It can be proved by the similar way as in Theorem 4.1 from \cite{DS}. 
Let us denote
\be\la{r2}
r_2(\ve,\tau)=\sum\limits_{\scriptsize \ba{c}m,n\in J,\,m\not=n\\
x\in I_{m N_\ve},\, y\in I_{n N_\ve}\ea}
{\cal G}^g_{\tau/{\ve}}(l+\!x)
Q_\ve([r/\ve]\!-\!x,[r/\ve]\!-\!y)({\cal G}^{g}_{\tau/{\ve}}(p+y))^T
\ee
and prove that
\be\la{r2(ve,t)}
r_2(\ve,\tau) \to 0\quad\mbox{as }\quad \ve\to0
\ee
 for any $\tau\in\R^1$.
We divide the sum in the RHS of (\ref{r2})
onto two sums $S_1$ and $S_2$, where
the first sum  $S_1$ is taken over all $x\in I_{m N_\ve}$, 
$y\in I_{n N_\ve}$
and $m,n\in J$ such that $\exists j\in\{1,\dots,d\}:|m_j-n_j|\ge2$;
the sum $S_2$ is taken over
all $x\in I_{m N_\ve}$, $y\in I_{n N_\ve}$
and $m,n\in J$ such that $m\not=n$ and
$\forall j=1,\dots,d: |m_j-n_j|\le1$.
The number of points $m\in J$ is order of $(\tau/(\ve N_\ve))^d$,
the number of points $x\in I_{mN_\ve}$ is $\sim N_\ve^d$.
Therefore, by Lemma  \ref{l5.3} (i) and condition {\bf V2},
the sum $S_1$ is estimated by
$$
C(1+\tau/{\ve})^{-d}(\tau/{\ve})^d
\sum\limits_{s\in\Z^d,|s|\ge N_\ve}(1+|s|)^{-\gamma},
%%\sim C_1\int\limits_{N_\ve}^{+\infty}e^{-a_0 r}r^{d-1}\,dr
$$
which vanishes as $\ve \to 0$, since $N_\ve\to+\infty$
as $\ve\to0$ and $\gamma >d$.
To estimate the second sum $S_2$ (the contribution
of nearest neighbors $I_{m N_\ve}$ and $I_{n N_\ve}$)
we choose a number $p>d+1$ and 
divide the sum onto two sums: $S_2=S_{21}+S_{22}$,
where the sum $S_{21}$ is taken over
all $m\in J$ and $x\in I_{mN_\ve}$,
$n\in\{n\in J:\, n\not=m, \forall j: |m_j-n_j|\le1\}$
and $y\in I_{nN_\ve}$ such that $|x-y|\ge N^{1/p}_{\ve}$
and the second sum $S_{22}$ is taken, respectively,
 over $y$ such that $|x-y|\le N^{1/p}_{\ve}$.
%%%%%%%%%%%%%%%%%%%%%%%%%%%%%%%%%%%%%%%%
The contribution of ``non-boundary zones'' $S_{21}$ is
$$
C(1+\tau/{\ve})^{-d}(\tau/{\ve})^d
\sum\limits_{s\in\Z^d,|s|\ge N^{1/p}_\ve}(1+|s|)^{-\gamma}
$$
which vanishes as $\ve \to 0$.
The contribution of ``boundary zones'' $S_{22}$
%$x'\in  A_{mN_\ve}$, $y'\in  A_{nN_\ve}$
%with $n\not= m$, s.t. $|n_j-m_j|\le 1$ for all $j=1,\dots,d$,
is order of
\be\la{3.23}
C(1+\tau/{\ve})^{-d}(\tau/{\ve} N_{\ve})^d N_{\ve}^{1/p+d-1}
N_{\ve}^{d/p}\sim C N_{\ve}^{(d+1)/p-1}.
\ee
%%%%%%%%%%%%%%%
Indeed, the number of points $m\in J$ is order of $(\tau/(\ve N_\ve))^d$,
the number of points $\{n: |m_j-n_j|\le1,\forall j, m\not=n\}$ is 
finite. For fixed $m,n$ the number of points $x\in I_{mN_\ve}$
such that $|x-y|\le N_\ve^{1/p}$ is order of $N_\ve^{d-1}N_{\ve}^{1/p}$.
For fixed $x$ the number of points $y$ such that $|x-y|\le N_\ve^{1/p}$ 
is $\sim N_\ve^{d/p}$.
The number $p$ is chosen  such that $(d+1)/p-1<0$. Hence,
 (\ref{3.23})  vanishes as $\ve \to 0$
by condition {\bf V1} (ii).
The decay (\ref{r2(ve,t)}) is proved.
%%%%%%%%%%%%%%%%%%%%%%%%%%%%%%%%%%%%

Therefore,
$$
S_{\ve,\tau/\ve}([r/\ve]\!+\!l,[r/\ve]+p)=
\sum\limits_{m\in J}
\sum\limits_{x, y\in I_{m N_\ve}}\!\!\!\!\!
{\cal G}^g_{\tau/\ve}(l+\!x)\bar Q_\ve([r/\ve]\!-\!x,[r/\ve]\!-\!y)
{\cal G}^{g}_{\tau/\ve}(p+y)^T+o(1).
$$
%%--------------------------------------------
{\it Step (iii)}
Now we can apply condition {\bf V1} (i) at the points
$[r/\ve]-x, [r/\ve]-y$ 
of the same cube $ I_{[r/\ve]-m N_\ve}$ and obtain
$$
|\bar Q_\ve([r/\ve]\!-\!x,[r/\ve]\!-\!y)-
\bar R(\ve[r/\ve]-\ve m N_\ve, [r/\ve]\!-\!x,[r/\ve]\!-\!y)|\le
C \min[(1+|x-y|)^{-\gamma},\ve N_{\ve}],
$$
where, by definition, the function $\bar R$ is equal to
$$
\bar R(r,x,y)=\left\{
\ba{cl} R(r,x,y)&\mbox{if }\, x,y\in\Z^d_+,\\
0& \mbox{otherwise} \ea\right.
$$
Therefore,
\beqn\la{3.24}
S_{\ve,\tau/\ve}([r/\ve]\!+\!l,[r/\ve]\!+\!p)\!\!&=&\!\!
\sum\limits_{m\in J}
\sum\limits_{x,y\in I_{m N_\ve}}\!\!\!
{\cal G}^g_{\tau/\ve}(l+\!x)
\bar R(\dots){\cal G}^{g}_{\tau/\ve}(p+y)^T+o(1),
\eeqn
where $\bar R(\dots)\equiv 
\bar R(\ve[r/\ve]-\ve m N_\ve, [r/\ve]-x,[r/\ve]-y)$.
Indeed, for fixed $x\in I_{mN_\ve}$, 
\beqn\nonumber
\sum\limits_{y\in I_{m N_\ve}}
\min[(1+|x-y|)^{-\gamma},\ve N_\ve]=
\sum\limits_{y:(1+|x-y|)^{-\gamma}\ge \ve N_\ve}\ve N_\ve
+ \sum\limits_{y:(1+|x-y|)^{-\gamma}\le \ve N_\ve}
(1+|x-y|)^{-\gamma}\nonumber\\
=\sum\limits_{s:(1+|s|)\le (N_\ve)^{-1/\gamma}}\ve N_\ve
+ \sum\limits_{s:(1+|s|)\ge (N_\ve)^{-1/\gamma}}
(1+|s|)^{-\gamma}\sim (\ve N_\ve)^{1-d/\gamma}.
\nonumber
\eeqn
%%%%%%%%%%%%%%%%%%%%%%%%%%%%%%%%%%%%
By Lemma  \ref{l5.3} (i) and condition {\bf V1} (ii), we obtain
\beqn
&&
\sum\limits_{m\in J}
\sum\limits_{x, y\in I_{m N_\ve}}
|{\cal G}^g_{\tau/{\ve}}(l+\!x)|\min[(1+|x-y|)^{-\gamma},\ve N_{\ve}]
|({\cal G}^{g}_{\tau/{\ve}}(p+y))^T|\nonumber\\
&\le& C(1+\tau/{\ve})^{-d}(\tau/(\ve N_\ve))^d N_\ve^d(\ve N_\ve)^{1-d/\gamma}
\sim \ve^{(1-\beta)(1-d/\gamma)} \to0,\,\,\,\,\ve\to0,
\nonumber
\eeqn
since $\beta<1$ and $\gamma>d$.
%%%%%%%%%%%%%%%%%%%%%%%%%%%%%%%%%%%%%

By the similar arguments as in steps (i) and (ii) of the proof,
the sums  in the RHS of (\ref{3.24}) 
can be taken  
over $\{m\in J,\,x\in I_{mN_\ve},\,y\in\Z^d\}$, i.e.,
$$
S_{\ve,\tau/\ve}([r/\ve]\!+\!l,[r/\ve]\!+\!p)=
\sum\limits_{m\in J}
\sum\limits_{x\in I_{m N_\ve}}\sum\limits_{y\in\Z^d}
{\cal G}^g_{\tau/\ve}(l+\!x)
\bar R(\dots){\cal G}^{g}_{\tau/\ve}(p+y)^T+o(1).
$$
%%%%%%%%%%%%%%%%%%%%%%%%%%%%%%%%%%%%%

{\it Step (iv)}
Let us split the function $\bar R$ into the following three matrix functions:
\beqn
R^+(r,x,y)&:=&\frac12 {\bf R}_0(r,x-y),\label{d1'}\\
R^-(r,x,y)&:=&\frac12 {\bf R}_0(r,x-y)\sign (y_1),\label{d1''}\\
R^0(r,x,y)&:=&\bar R(r,x,y)-R^+(r,x,y)-R^-(r,x,y).\label{d1'''}
\eeqn
Next, introduce the  matrices
\be\label{Qta}
S^a_{\ve,\tau/\ve}\equiv S^a_{\ve,\tau/\ve} ([r/\ve]\!+\!l,[r/\ve]\!+\!p)=
\sum\limits_{m\in J}
\sum\limits_{x\in I_{m N_\ve}}\sum\limits_{y\in\Z^d}
{\cal G}^g_{\tau/\ve}(l+\!x)
\bar R^a(\dots){\cal G}^{g}_{\tau/\ve}(p+y)^T,
\ee
for each $a=\{+,-,0\}$ and split 
$S_{\ve,\tau/\ve}$ into three terms,
$S_{\ve,\tau/\ve}=S^+_{\ve,\tau/\ve}+S^-_{\ve,\tau/\ve}+S^0_{\ve,\tau/\ve}$.
The convergence (\ref{pro4}) results now from the following three lemmas. 
%%------------------------------------------
\begin{lemma}\label{Qt1}
 $\lim\limits_{\ve\to0} S^+_{\ve,\tau/\ve}([r/\ve]\!+\!l,[r/\ve]\!+\!p)
= \frac12 s^{+}_{\tau,r}(l-p)$, $l,p\in\Z^d$, 
where $s^{+}_{\tau,r}$ is defined as in (\ref{qtaur})--(\ref{Pi}) but with
$
\ds\frac{1}2
\Big(\hat {\bf R}_0(r+\nabla\om_\sigma(\theta)\tau,\theta)
\pm\hat {\bf R}_0(r-\nabla\om_\sigma(\theta)\tau,\theta)\Big)
$
instead of ${\bf R}^{\sigma}_\pm$ (cf. (\ref{gclimcor2})).
\end{lemma}
%%----------------------
{\bf Proof.} 
By (\ref{d1'}) and (\ref{Qta}), the function $S^+_{\ve,\tau/\ve}$
can be represented as
$$
S^+_{\ve,\tau/\ve}=
\frac12\sum\limits_{m\in J}\sum\limits_{x\in I_{m N_\ve}}
{\cal G}^g_{\tau/\ve}(l+x)\sum\limits_{y\in \Z^d}
 {\bf R}_0(\ve[r/\ve]-\ve m N_\ve, y-x){\cal G}^g_{\tau/\ve}(p+y)^T.
$$
Using Fourier transform and the Parseval equality
we can rewrite $S^+_{\ve,\tau/\ve}$ as
\beqn\label{4.9}
S^+_{\ve,\tau/\ve}&=&(2\pi)^{-2d}\frac12
\sum\limits_{m\in J} \sum\limits_{x\in I_{m N_\ve}}
\int\limits_{\T^{2d}}
e^{-i(l\cdot\theta-p\cdot\theta')}e^{i x\cdot(\theta'-\theta)}
\hat{\cal G}^g_{\tau/\ve}(\theta)
\hat {\bf R}_0(\ve[r/\ve]-\ve m N_\ve,\theta')
\nonumber\\
&&\times\hat{\cal G}^g_{\tau/\ve}(\theta')^*\,d\theta'd\theta.
\eeqn
Therefore, the proof of  Lemma \ref{Qt1} reduces to the 
finding the limit value of (\ref{4.9}), that is done in
Theorem 4.1 from \cite{DS} (the detailed proof see in Appendix~A). 
%%---------------------------------------
\begin{lemma}\label{Qt2}
 $\lim\limits_{\ve\to0}
S^-_{\ve,\tau/\ve}([r/\ve]\!+\!l,[r/\ve]\!+\!p)
= \frac12 s^{-}_{\tau,r}(l-p) $, $l,p\in\Z^d$, 
where $s^{-}_{\tau,r}$ is defined as in (\ref{qtaur})--(\ref{Pi}) but with
$$
\frac{1}2
\Big(\hat {\bf R}_0(r+\nabla\om_\sigma(\theta)\tau,\theta)
\sign(r_1+\tau\nabla_1\om_\sigma(\theta))
\pm\hat {\bf R}_0(r-\nabla\om_\sigma(\theta)\tau,\theta)
\sign(r_1-\tau\nabla_1\om_\sigma(\theta))\Big)
$$
instead of ${\bf R}^{\sigma}_\pm$.
\end{lemma}
%%---------------------------------------
\begin{lemma}\label{Qt3}
 $\lim\limits_{\ve\to0} 
S^0_{\ve,\tau/\ve}([r/\ve]\!+\!l,[r/\ve]\!+\!p)=0$, $l,p\in\Z^d$. 
\end{lemma}
%%------------------------------------------------
The proofs of Lemmas \ref{Qt2} and \ref{Qt3} see in Appendices~B and C, resp.

%%%%%%%%%%%%%%%%%%%%%%%%%%%%%%%%%%%%%%%%%%%%%%%%%%
%%%%%%%%%%%%%%%%   section 4 %%%%%%%%%%%%%%%%%%
%%%%%%%%%%%%%%%%%%%%%%%%%%%%%%%%%%%%%%%%%%%%%%%%%%
\setcounter{equation}{0}
\section{Convergence of Wigner matrices}\label{sec.4}
%%%%%%%%%%%%%%%%%%%%%%%%%%%%%%%%%%%%%%%%%%%
Here we prove Theorem \ref{the2}.
Theorem \ref{the2'} implies that
 for any $r\in\R^d_+$,
$\tau\not=0$ and $y\in (2\Z)^d$, the following convergence holds,
\beqn\la{5.1}
\lim_{\ve\to0}
E^\varepsilon_{\tau/\varepsilon}
\big(a([r/\ve]+y/2)^\ast \otimes a([r/\ve]-y/2)\big)
={\cal W}^p(\tau;r,y),
\eeqn
where in the Fourier space one has
\beqn\la{5.2}
\hat {\cal W}^p(\tau;r,\theta)&=&
 \frac{1}{2}\Big(\Omega^{1/2}\hat q^{G,00}_{\tau,r}(\theta)\Omega^{1/2}+
 \Omega^{-1/2}\hat q^{G,11}_{\tau,r}(\theta)\Omega^{-1/2}
\nonumber\\&&
+i\Omega^{1/2}\hat q^{G,01}_{\tau,r}(\theta)\Omega^{-1/2}-i
 \Omega^{-1/2}\hat q^{G,10}_{\tau,r}(\theta)\Omega^{1/2}\Big)
\nonumber\\
&=&W^p(\tau;r,\theta),
\eeqn
by formulas (\ref{3.16a}) and (\ref{qtaur})--(\ref{gclimcor2}).
Therefore, convergence (\ref{3.17}) follows from (\ref{5.1}),
(\ref{5.2}) and Lemma \ref{lcom}.
%%%%%%%%%%%%%%%%%%%%%%%%%%%%%%%
\begin{lemma}\label{lcom}
Let conditions {\bf V2}  and {\bf E1}--{\bf E3}, {\bf E6} hold
and let $\alpha<-d/2$. Then the following bound holds:
$$
\sup\limits_{t\in\R} \sup\limits_{z,z'\in \Z^d_+}
\Vert Q_{\ve,t}(z,z')\Vert\le C<\infty.
$$
\end{lemma}
%%%%%%%%%%%%%%%%%%%%%%%%%%%%%%%%%
{\bf Proof}\,
The representation (\ref{sol}) gives
\beqn
Q^{ij}_{\ve,t}(z,z')&=&E^\ve_0\Big(Y^i(z,t)\otimes Y^j(z',t)\Big)
= \sum\limits_{y,y'\in \Z^d_+} \sum\limits_{k,l=0,1}
{\cal G}^{ik}_{t,+}(z,y)Q^{kl}_\ve(y,y'){\cal G}^{jl}_{t,+}(z',y')\nonumber\\
&=& \langle Q_\ve(y,y'), \Phi^i_{z}(y,t)\otimes
\Phi^j_{z'}(y',t)\rangle_+,\nonumber
\eeqn
where $\Phi^i_{z}(y,t)$ is given by
\beqn
\Phi^i_{z}(y,t)&=&\Big(
{\cal G}^{i0}_{t,+}(z,y),{\cal G}^{i1}_{t,+}(z,y)\Big)\nonumber\\
&=&({\cal G}_t^{i0}(z-y)-{\cal G}_t^{i0}(z-\tilde y),
{\cal G}_t^{i1}(z-y)-{\cal G}_t^{i1}(z-\tilde y)),
\,\,\,\,\,i=0,1.\nonumber
\eeqn
The Parseval identity, formula (\ref{4.5}),
and condition {\bf E6} imply that
$$
\Vert\Phi^i_{z}(\cdot,t)\Vert^2_{l^2}= (2\pi)^{-d}
\int\limits_{\T^d} |\hat\Phi^i_{z}(\theta,t)|^2\,d\theta
\le C\int\limits_{\T^d}
\Big( |\hat{\cal G}^{i0}_t(\theta)|^2
+|\hat{\cal G}^{i1}_t(\theta)|^2\Big)
\,d\theta \le C_0<\infty.
$$
Corollary \ref{c4.10} gives now
$$
|Q^{ij}_{\ve,t}(z,z')|=
|\langle Q_\ve(y,y'), \Phi^i_{z}(y,t)\otimes
\Phi^j_{z'}(y',t)\rangle_+|
\le C\Vert\Phi^i_{z}(\cdot,t)\Vert_{l^2_+}\,
\Vert\Phi^j_{z'}(\cdot,t)\Vert_{l^2_+}\le C_1<\infty,
$$
where the constant $ C_1$  does  not depend on
$z,z'\in\Z^d_+$, $t\in\R$, and $\ve>0$.\bo

%%%%%%%%%%%%%%%%%%%%%%%%%%%%%%%%%   5  %%%%%%%%%%%%%%%%%%%%%
\setcounter{equation}{0}
\section{Proof of Theorem 2.14}\label{sec.5}
%%%%%%%%%%%%%%%%%%%%%%%%%%%%%%%%%%%%%%%%%%%%%%
Theorem  \ref{the3} follows  from Propositions  \re{l2.1} and \re{l2.2}.
Proposition \ref{l2.1} ensures the existence of the limit measures
of the family $\{\mu^\ve_{\tau/\ve,r},\,\ve>0\}$,
while Proposition \ref{l2.2} provides the
 uniqueness.
%%%-------------------------------
\begin{pro}\la{l2.1}
Let conditions {\bf V2}  and {\bf E1}--{\bf E3}, {\bf E6} hold.
Then for any $r\in \R^d$ with $r_1\ge0$, $\tau\not=0$, the family of
measures $\{\mu^\ve_{\tau/\ve,r},\,\ve>0\}$
 is weakly compact on ${\cal H}_{\al,+}$ for any
 $\al<-d/2$, and the following bounds hold:
\be\la{p3.1}
\sup\limits_{\ve\ge 0}
 \int \Vert Y_0\Vert^2_\al\,\mu^\ve_{\tau/\ve,r}(dY_0) <\infty.
\ee
 \end{pro}
%%%%%%%%%%%%%%%%%%%%%%%%%%
{\bf Proof.}\,
Definition \ref{d1.1} yields
\beqn\nonumber
&& \int\Vert  Y_0\Vert^2_{\alpha,_+}\mu^{\ve}_{\tau/\ve,r}(dY_0)=
E^\ve_0 \big(\Vert T_{-[r/\ve]} U_+(\tau/\ve)Y_0\Vert^2_{\al,+}\big)
\nonumber\\
&=&\sum\limits_{z\in \Z^d_+} (1+|z|^2)^\alpha
{\rm tr}\Big(Q_{\ve,\tau/\ve}^{00}([r/\ve]+z,[r/\ve]+z)+
Q_{\ve,\tau/\ve}^{11}([r/\ve]+z,[r/\ve]+z)\Big).\la{5.2estimate}
\eeqn
Since  $\alpha<-d/2$, estimate (\ref{p3.1}) follows from Lemma \ref{lcom}
and (\ref{5.2estimate}).
Now the compactness of the
measures  family  $\{\mu^\ve_{\tau/\ve,r},\,\ve>0\}$ follows
from the bound (\re{p3.1})
by the Prokhorov compactness theorem  \cite[Lemma II.3.1]{VF}
by using a method applied in \ci[Theorem XII.5.2]{VF}
because the embedding ${\cal H}_{\al,+} \subset{\cal H}_{\beta,+}$ 
is compact for $\al>\beta$.
\bo\medskip

%%%%%%%%%%%%%%%%%%%%%%%%%%%%%%%%%%%%%
Set  ${\cal S}_+=[S(\Z_+^d)\times \R^n]^2$,
where $S(\Z^d_+)$ stands for the space of rapidly decreasing real sequences
on $\Z^d_+$. 
Write $\langle Y,\Psi\rangle_+=\langle Y^0,\Psi^0\rangle_++
\langle Y^1,\Psi^1\rangle_+$ for $Y=(Y^0,Y^1)\in{\cal H}_{\al,+}$
and $\Psi=(\Psi^0,\Psi^1)\in {\cal S}_+$,
where $\langle Y^i,\Psi^i\rangle_+=\sum_{z\in\Z^d_+}
Y^i(z)\cdot\Psi^i(z)$, $i=0,1$.
%%%%%%%%%%%%%%%%%%%%%%%%%%%%%%%%%%%%%%%%%%5
 \begin{pro}\la{l2.2}
Let conditions {\bf V1}--{\bf V4} and {\bf E1}--{\bf E6} hold.
Then for any $r\in \R^d$ with $r_1\ge0$, $\tau\not=0$
and $\Psi\in {\cal S}_+$,
the characteristic functionals converge to a Gaussian one,
 \be\la{2.6}
\hat \mu^\ve_{\tau/\ve,r}(\Psi):=
 \int e^{i\langle Y,\Psi\rangle_+}\mu^\ve_{\tau/\ve,r}(dY)
\to \exp\big\{-\fr{1}{2}{\cal Q}_{\tau,r} (\Psi,\Psi)\big\}
=:\hat \mu^G_{\tau,r}(\Psi)\,\,\,
as \,\,\,\ve\to0,
 \ee
where
${\cal Q}_{\tau,r}$ is the quadratic form with the matrix  kernel
$(Q^{G}_{\tau,r}(x,y))_{i,j=0,1}$,
\be\la{qpp'}
{\cal Q}_{\tau,r}(\Psi,\Psi)=\sum\limits_{i,j=0,1}~
\sum\limits_{z,z'\in\Z^d_+}
\big( Q^{G,ij}_{\tau,r}(z,z'),\Psi^i(z)\otimes\Psi^j(z')\big),
\,\,\,\,\Psi\in{\cal S}_+.
\ee
 \end{pro}
%%%---------------------------------

To prove Theorem \ref{the3} it remains to check Proposition \ref{l2.2}.
Let us rewrite (\re{2.6}) as
\be\la{2.6i}
\hat \mu^\ve_{\tau/\ve,r}(\Psi)= E^\ve_0
\big(\exp \{i\langle T_{-[r/\ve]}U_+(\tau/\ve)Y_0,\Psi\rangle_+\}\big)
\to \hat \mu^G_{\tau,r}(\Psi),\,\,\,\, \ve\to0.
\ee
We derive (\ref{2.6i}) by using the explicit representation (\ref{sol}) 
of the solution $Y(t)$, the Bernstein `room--corridor' technique
and the approach of \cite{DS,D08}. The approach gives a representation 
of $\langle T_{-[r/\ve]}U_+(\tau/\ve)Y_0,\Psi\rangle_+$ 
as a sum of weakly dependent random variables 
(see formula (\ref{razli}) below). 
In this case, (\ref{2.6i})  follows from the central 
limit theorem under a Lindeberg-type condition.
%A similar technique of proof was applied in \cite{DS, D08}.
%Therefore, we represent only the main steps of the proof.

%%%%%%%%%%%%%%%%%%   5.1   %%%%%%%%%%%%
\subsection{Duality arguments}
%%%%%%%%%%%%%%%%%%%%%%%%%%%%%%%%%%%%
In   this section, we evaluate the inner product
$\langle T_{-[r/\ve]}U_+(\tau/\ve)Y_0,\Psi\rangle_+$.
Introduce the function $\Psi_*(z)$ as
$\Psi_*(z)=\Psi(z)$ for $z_1\ge 0$, and $\Psi_*(z)=0$ otherwise.
Therefore, 
\beqn\label{YP}
\langle T_{-[r/\ve]}U_+(\tau/\ve)Y_0,\Psi(z)\rangle_+=
\langle Y_0(z'),\Phi_r(z',\tau/\ve)\rangle_+,
\eeqn
where, by definition, the function $\Phi_r(z',\tau/\ve)$ is equal to 
\beqn\label{7.2}
\Phi_r(z',\tau/\ve)&=&\sum\limits_{z\in\Z^d_+} 
{\cal G}_{\tau/\ve,+}^{T}(z,z')T_{[r/\ve]}\Psi_*(z)
\nonumber\\
&=&(2\pi)^{-d}\int\limits_{\T^d} 
(e^{-iz'\cdot\theta}-e^{-i\ti z'\cdot\theta})\hat{\cal G}^*_t
(\theta) e^{i[r/\ve]\cdot \theta}\hat\Psi_*(\theta)\,d\theta.
\eeqn
%%%%%%%%%%%%%%%%%%%%%%%%%%%%%%%%%%%%%%%%%%%%%%%%%%%%%%%%%%%%
Let us denote
\be\la{calS}
{\cal S}^0=\{\Psi\in{\cal S}=[S(\Z^d)\otimes\R^n]^2:
\hat\Psi(\theta)=0\,\, \mbox{\rm in a neighborhood of}\,\,{\cal C}\},
\ee
where ${\cal C}$ is defined in (\ref{calC}).
Since {\rm mes}\,${\cal C}=0$ 
it suffices to prove (\ref{2.6i}) for $\Psi_*\in {\cal S}^0$ only. 
For the function $\Phi_r(z,\tau/\ve)$, the following lemma holds.
%%------------------------------------------
\begin{lemma}\label{l5.3'} (cf. Lemma 6.3 from \cite{D08}, 
Lemma 5.2 from \cite{DS}).
Let conditions {\bf E1}--{\bf E4} and {\bf E6} hold. Then,
for any chosen $\Psi_*\in {\cal S}^0$,  the following bounds hold.\\
 (i) $\sup_{z\in\Z^d}|\Phi_r(z,\tau/\ve)| \le  C~\ve^{d/2}$.\\
(ii) For  any $p>0$, there exist $C_p>0$ and $\gamma=\gamma(\tau,r)>0$ 
such that 
\be\label{conp}
|\Phi_r(z,\tau/\ve)|\le C_p(1+|z|+\tau/\ve)^{-p},
\quad |z|\ge\gamma\tau/\ve.
\ee
\end{lemma}

This lemma follows from (\ref{7.2}), (\ref{calS}), (\ref{4.5}), 
and the standard stationary phase method.

%%%%%%%%%%%%%%%%%%%%%  5.2   %%%%%%%%%%%%
\subsection{Bernstein's room-corridor' partition}
%%%%%%%%%%%%%%%%%%%%%%%%%%%%%%%%%%%%%%%%%%%%%%%%%%%%%%%%%%
Write $t=\tau/\ve$.
Let us  introduce a `room--corridor'  partition of the
half-ball $\{z\in\Z^d_+:~|z|\le \gamma t\}$  with $\gamma$ in (\ref{conp}).
For $t>0$, we choose $\Delta_t$ and $\rho_t\in\N$.
Choose a $\delta$, $0<\delta<1$, and
\be\label{rN}
\rho_t\sim t^{1-\delta},
~~~\Delta_t\sim\frac t{\log t},~~~~\,\,\,t\to\infty.
\ee
Write $h_t=\Delta_t+\rho_t$ and   
$a^j=jh_t$, $b^j=a^j+\Delta_t$, $j=0,1,2,\dots$, $n_t=[(\gamma t)/h_t]$.   
We refer to the slabs $R_t^j=\{z\in\Z^d_+:|z|\le n_t h_t,\,a^j\le z_1< b^j\}$ 
as the `rooms',   to
$C_t^j=\{z\in\Z^d_+: |z|\le n_t h_t,\, b^j\le z_1<a^{j+1}\}$ as
the `corridors', and to
$L_t=\{z\in\Z^d_+: |z|> n_t h_t\}$ as the 'tail'.   
Here  $z=(z_1,\dots,z_d)$, $\Delta_t$ stands for the width of the room, and   
$\rho_t$ for that of the corridor. Denote  by   
 $\chi_t^j$ the indicator of the room $R_t^j$, 
by $\xi_t^j$ that of the corridor $C_t^j$, and  
by $\eta_t$ that of the tail $L_t$. In this case,  
$$
{\sum}_j [\chi_t^j(z)+\xi_t^j(z)]+ \eta_t(z)=1,\,\,\,z\in\Z^d_+,  
$$   
where the symbol ${\sum}_j$ stands for the sum $\sum\limits_{j=0}^{n_t-1}$.   
Hence, we obtain the following  Bernstein-type representation:   
\be\label{res}   
\langle Y_0,\Phi_r(\cdot,t)\rangle_+ = {\sum}_j   
\left[\langle Y_0,\chi_t^j\Phi_r(\cdot,t)\rangle_+ +   
\langle Y_0,\xi_t^j\Phi_r(\cdot,t)\rangle_+ \right]+   
\langle Y_0,\eta_t\Phi_r(\cdot,t)\rangle_+.   
\ee   
Introduce the random variables $r_{t}^j$, $c_{t}^j$, $l_{t}$ by   
$$
r_{t}^j= \langle Y_0,\chi_t^j\Phi_r(\cdot,t)\rangle_+,~~   
c_{t}^j= \langle Y_0,\xi_t^j\Phi_r(\cdot,t)\rangle_+,   
\,\,\,l_{t}= \langle Y_0,\eta_t\Phi_r(\cdot,t)\rangle_+.   
$$   
Therefore, it follows from (\ref{YP}) and (\ref{res})  that 
\be\label{razli}   
\langle T_{-[r/\ve]}U_+(t)Y_0,\Psi\rangle_+
=\langle Y_0,\Phi_r(\cdot,t)\rangle_+ =   
{\sum}_j (r_{t}^j+c_{t}^j)+l_{t}.   
\ee   
%%%%%%%-------------------------------------------------
\begin{lemma}  \label{l5.1} 
    Let conditions {\bf V1} and {\bf V2} hold and $\Psi_*\in{\cal  S}^0$.
The following bounds hold for $t>1$:
\beqn
E_0^\ve|r^j_{t}|^2&\le&  C(\Psi)~\Delta_t/ t,\,\,\,\forall j,\la{106}\\
E_0^\ve|c^j_{t}|^2&\le& C(\Psi)~\rho_t/ t,\,\,\,\forall j,\la{106''}\\
E_0^\ve|l_{t}|^2&\le& C_p(\Psi)~t^{-p},\,\,\,\,\forall p>0.\la{106'''}
\eeqn
\end{lemma}

The proof is based on Lemmas \ref{l4.1} and \ref{l5.3'} 
(see also \cite[Lemma 7.1]{DS}).
%%%%%%%%%%%%%%%%%%%%%%%%%%%%%%%%%%%%%%%%%%%%%%%%%%%%%%%%
\medskip

Further, to prove (\ref{2.6i}), we use a version of the central limit theorem
developed by Ibragimov and Linnik.
If  ${\cal Q}_{\tau,r}(\Psi,\Psi)=0$, then the convergence (\ref{2.6}) is 
obvious. Indeed, then,
\beqn\nonumber
&&\left| E^\ve_0\big(
\exp\{i \langle Y_0,\Phi_r(\cdot,\tau/\ve)\rangle_+\}\big) -
 \hat \mu^G_{\tau,r}(\Psi)\right|
= E^\ve_0\big(
|\exp\{i\langle Y_0,\Phi_r(\cdot,\tau/\ve)\rangle_+\}-1|\big)
\nonumber\\
&\le& E^\ve_0\big(|\langle Y_0,\Phi_r(\cdot,\tau/\ve)\rangle_+|\big)
\le
\left(E^\ve_0\big(|\langle Y_0,\Phi_r(\cdot,\tau/\ve)\rangle_+|^2
\big)\right)^{1/2}
\nonumber\\
&=&\big(\langle Q_\ve(x,y),\Phi_r(x,\tau/\ve)\otimes
\Phi_r(y,\tau/\ve)\rangle_+\big)^{1/2}
=\big({\cal Q}_{\ve,\tau/\ve,r}(\Psi,\Psi)\big)^{1/2},
\eeqn
where ${\cal Q}_{\ve,\tau/\ve,r}(\Psi,\Psi)\to
{\cal Q}_{\tau,r}(\Psi,\Psi)=0$, $\ve\to0$.
Therefore,  (\ref{2.6}) follows from Theorem \ref{the2'}.

Thus, we may assume that, for a given $\Psi_*\in{\cal S}^0$,
\be\label{5.*}
{\cal Q}_{\tau,r}(\Psi,\Psi)\not=0.
\ee
%%%%%%%%%%%%%%%%%%%%%%%%%%%%%%%%%%%%5
\begin{lemma}\la{r}
The following limit holds,
\be\la{7.15'}
n_t\Big[\Bigl(\frac{\rho_t}{t}\Bigr)^{1/2}+
(1+\rho_t)^{-\gamma}\Big]+
n_t^2\frac{\rho_t}{t}\to 0 ,\quad t\to\infty.
\ee
\end{lemma}
Indeed,  (\ref{rN}) implies that
$h_t=\rho_t+\De_t\sim \ds\frac{t}{\log t}$,
 $t\to\infty$.
Therefore, $n_t\sim\ds\frac{t}{h_t}\sim\log t$.
Then  (\ref{7.15'}) follows by  (\ref{rN}).
\bo\\

For simplicity, we put $t =\tau/\ve$.
By the triangle inequality,
\beqn
\Big|E^\ve_0\big(\exp\{i \langle Y_0,\Phi_r(\cdot,t)\rangle_+\}
\big)
-\hat \mu^G_{\tau,r}(\Psi)\Big|
&\le &
\Big|E^\ve_0\big(\exp\{i \langle Y_0,\Phi_r(\cdot,t)\rangle_+\}\big)
-E^\ve_0\big(\exp\{i{\sum}_j r_{t}^j\}\big)\Big|
\nonumber\\
&&\hspace{-12pt}+\Big|\exp\big\{-\frac{1}{2}{\sum}_j 
E^\ve_0\big(|r_{t}^j|^2\big)\big\}
 \!-\!
\exp\big\{-\frac{1}{2} {\cal Q}_{\tau,r}(\Psi, \Psi)\big\}\Big|
 \nonumber\\
&&\hspace{-12pt}+ \Big|E^\ve_0 \big(\exp\{i{\sum}_j r_{t}^j\}\big) \!-\!
\exp\big\{-\frac{1}{2}{\sum}_j 
E^\ve_0\big(|r_{t}^j|^2\big)\big\}\Big|\nonumber\\
&=& I_1+I_2+I_3. \la{4.99}
\eeqn
We are going to   show  that all summands
$I_1$, $I_2$, $I_3$  tend to zero
 as  $t\to\infty$.\\
{\it Step (i)} Eqn (\ref{razli}) implies
\beqn\la{101}
I_1&=&\Big|E^\ve_0\big(\exp\{i{\sum}_j r^j_{t} \}
\big(\exp\{i{\sum}_j c^j_{t}+il_{t}\}-1\big)\big)\Big|
\nonumber\\
&\le&
 {\sum}_j E^\ve_0\big(|c^j_t|\big)+
E^\ve_0\big(|l_{t}|\big)
\le{\sum}_j\Big(E^\ve_0\big(|c^j_t|^2\big)\Big)^{1/2}+
\Big(E^\ve_0\big(|l_{t}|^2\big)\Big)^{1/2}.
\eeqn
From (\ref{101}), (\ref{106''}), (\ref{106'''})
  and (\ref{7.15'}) we obtain that
\beqn\nonumber%\la{103}
I_1\le C n_t(\rho_t/t)^{1/2}+C_p t^{-p} \to 0,~~t\to \infty.
\eeqn
%%%%%%%%%%%%%%%%%%%%%%%%%%%%%%%%%%%%%%%%%%%%%%%%%%%%
{\it Step (ii)} By the triangle inequality,
\beqn
I_2&\le& \frac{1}{2}
\Big|{\sum}_j E^\ve_0(|r_t^j|^2)- 
{\cal Q}_{\tau,r}(\Psi,\Psi)\Big|
\le \frac{1}{2}\,
\Big|{\cal Q}_{\ve,t,r}(\Psi, \Psi)-
{\cal Q}_{\tau,r}(\Psi,\Psi)\Big|
\nonumber\\
&&+ \frac{1}{2}\, \Big|E^\ve_0\big(\big({\sum}_j r_t^j\big)^2\big)
-{\sum}_j E^\ve_0\big(|r_t^j|^2\big)\Big| +
 \frac{1}{2}\, \Big|E^\ve_0\big(\big({\sum}_j r_t^j\big)^2\big)
-{\cal Q}_{\ve,t,r}(\Psi, \Psi)\Big|\nonumber\\
&=& I_{21} +I_{22}+I_{23}\la{104},
\eeqn
where ${\cal Q}_{\ve,t,r}$ is the quadratic form with
the  matrix kernel $Q_{\ve,t,r}^{ij}(x,y)$.
Theorem \ref{the2} implies that  $I_{21}\big|_{t=\tau/\ve}\to 0$
as $\ve\to0$.
As for  $I_{22}$,  we first obtain that
\be\la{i22}
I_{22}\le \sum\limits_{j<l} 
\left|E^\ve_0 \big(r_t^j r_t^l\big)\right|.
\ee
The distance between the different rooms $R_t^j$
is greater or equal to $\rho_t$.
Then, by Lemma \ref{l5.3'} (i) and condition {\bf V2},
\beqn\la{i222}
I_{22}&\le&\sum\limits_{j< l}
| \langle Q_\ve(x,y),\chi_t^j\Phi_r(x,t)\otimes
\chi_t^l\Phi_r(y,t)\rangle_+|\nonumber\\
&\le&Ct^{-d}\sum\limits_{j< l}
\sum\limits_x \chi_t^j(x)
\sum\limits_y\chi_t^l(y) (1+|x-y|)^{-\gamma}
\nonumber\\
&\sim&
 t^{-d}n_t^2 t^{d-1}\Delta_t
\int\limits_{\rho_t}^{+\infty}(1+s)^{-\gamma}s^{d-1}\,ds
\sim n_t (1+\rho_t)^{-\gamma+d},
\eeqn
which vanishes as $t\to\infty$ because of
 (\ref{7.15'}) and $\gamma>d$.
Finally, it remains to check  that $I_{23}\to 0$,
$t\to\infty$. We have
$$
{\cal Q}_{\ve,t,r}(\Psi,\Psi)
=E^\ve_0\big(\langle Y_0,\Phi_r(\cdot,t)\rangle_+^2\big)
=E^\ve_0\Big(\Big({\sum}_j (r_t^j+c_t^j)+l_t\Big)^2\Big),
$$
according  to (\ref{razli}).
Therefore, by the Cauchy-Schwarz inequality,
\beqn
I_{23}&\le& \Big|
E^\ve_0\big(\bigl({\sum}_j r_t^j\bigr)^2\big)
- E^\ve_0\big(\bigl({\sum}_j r_t^j +
{\sum}_j c_t^j+l_t\bigr)^2 \big)\Big|\nonumber\\
& \le&
C n_t{\sum}_j E^\ve_0 \big(|c_t^j|^2\big)  +
C_1\Bigl(
E^\ve_0\big(({\sum}_j r_t^j)^2\big)\Bigr)^{1/2}
\Bigl(
n_t{\sum}_j E^\ve_0\big(|c_t^j|^2\big)+
E^\ve_0 \big(|l_t|^2\big)\Bigr)^{1/2}\nonumber\\
&&+C  E^\ve_0 \big(|l_t|^2\big).\la{107}
\eeqn
Then  (\ref{106}), (\ref{i22}) and (\ref{i222}) imply
\beqn
E^\ve_0\big(({\sum}_j r_t^j)^2\big)&\le&
{\sum}_j E^\ve_0\big(|r_t^j|^2\big)
 \!+\!2{\sum}_{j<l}\Big|E^\ve_0\big( r_t^j r_t^l\big)\Big|
\nonumber\\
&\le&
Cn_t\De_t/t+C_1n_t(1+\rho_t)^{-\gamma+d}\le C_2<\infty.
\nonumber
\eeqn
Now (\ref{106''}), (\ref{106'''}), (\ref{107}), and (\ref{7.15'}) yield
\beqn\nonumber%\la{106'}
I_{23}\le C_1  n_t^2\rho_t/t+C_2 n_t(\rho_t/t)^{1/2}
+C_3 t^{-p} \to 0,~~t\to \infty.
\eeqn
So,  the terms $I_{21}$, $I_{22}$, $I_{23}$
in  (\ref{104}) tend to zero.
Then  (\ref{104}) implies that for $t=\tau/\ve$
\be\la{108}
I_2\le \frac{1}{2}\,
\left|{\sum}_j E^\ve_0\big(|r_t^j|^2\big)-
 {\cal Q}_{\tau,r}(\Psi, \Psi)\right|
\to 0,~~\ve\to0.
\ee
%%%%%%%%%%%%%%%%%%%%%%%%5
{\it Step (iii)}
It remains to verify that for $t=\tau/\ve$
$$
I_3=\Big|E^\ve_0\big(\exp\big\{i{\sum}_j r_t^j\big\}\big)
-\exp\big\{
-\fr12 {\sum}_j E^\ve_0\big(|r_t^j|^2\big)\big\}\Big|
 \to 0,~~\ve\to0.
$$
Condition {\bf V3} yields
\beqn
&&\Big|E^\ve_0\big(\exp\{i{\sum}_j r_t^j\}\big)
-\prod\limits_{-n_t}^{n_t-1}
E^\ve_0\big(\exp\{i r_t^j\}\big)\Big|\nonumber\\
&\le&
\Big|E^\ve_0\big(\exp\{ir_t^{-n_t}\}
\exp\{i\sum\limits_{-n_t+1}^{n_t-1} r_t^j\}\big)  -
 E^\ve_0\big(\exp\{ir_t^{-n_t}\}\big)E^\ve_0
\big(\exp\{i\sum\limits_{-n_t+1}^{n_t-1} r_t^j\}\big)\Big|
\nonumber\\
&&+
\Big|E^\ve_0\big(\exp\{ir_t^{-n_t}\}\big)
E^\ve_0\big(\exp\big\{i\sum\limits_{-n_t+1}^{n_t-1} r_t^j\big\}\big)
-\prod\limits_{-n_t}^{n_t-1}
E^\ve_0\big(\exp\{i r_t^j\}\big)\Big|
\nonumber\\
&\le& C(1+\rho_t)^{-2\gamma}+
\Big|E^\ve_0\big(\exp\big\{i\sum\limits_{-n_t+1}^{n_t-1} r_t^j\big\}\big)
-\prod\limits_{-n_t+1}^{n_t-1}
E^\ve_0\big(\exp\{i r_t^j\}\big)\Big|.\nonumber
\eeqn
We then apply condition {\bf V3} recursively
and obtain, according to Lemma \ref{r},
$$
\Big|E^\ve_0\big(\exp\{i{\sum}_j r_t^j\}\big)-
\prod\limits_{-n_t}^{n_t-1}
E^\ve_0\big(\exp\{i r_t^j\}\big)\Big|_{t=\tau/\ve}
\le
C n_{\tau/\ve}(1+\rho_{\tau/\ve})^{-2\gamma}\to 0,\quad \ve\to0.
$$
%%%%%%%%%%%%%%%%%%%%%%%%%%%%%%%%%%%%%
%We first obtain for $t=\tau/\ve$,
%$$
%| E_0^\ve\exp\{i \langle Y_0,\Phi_r(\cdot,t)\rangle_+\}- \hat\mu^G_{\tau,r}(\Psi)|
%=\left|E_0^\ve\exp\left\{i{\sum}_j r_t^j\right\}
%-\exp\left\{-\frac12 {\sum}_j E_0^\ve|r_t^j|^2\right\}\right|+o(1),\,\,\ve\to0.
%$$
%This follows from Lemma \ref{l5.1}, from the convergence (\ref{gclimcor0}),
%from condition {\bf V3}, and from the bound (\ref{rN}) (cf \cite{DS}).
%Next, by the mixing condition {\bf V3}, we see that
%$$
%\left|E_0^\ve\exp\left\{i{\sum}_j r_t^j\right\}-\prod\limits_{0}^{n_t-1}
%E_0^\ve\exp\left\{i r_t^j\right\}\right|
%\le C n_t\varphi_\ve(\rho_t)\big|_{t=\tau/\ve}\to 0,\quad \ve\to0.
%$$
%%%%%%%%%%%%%%%%%%%%%%%%%%%%%%555
Hence, it remains to show that for $t=\tau/\ve$
$$
\left|\prod\limits_{0}^{n_t-1} E_0^\ve\exp\left\{ir_t^j\right\}
-\exp\left\{-\frac12{\sum}_{j} E_0^\ve|r_t^j|^2\right\}
\right| \to 0,~~\ve\to0.
$$
According to the standard statement of the central limit theorem
(see, e.g. \cite[Theorem 4.7]{P}), 
it suffices to verify the  Lindeberg condition:
$$
\forall\delta>0,\,\,\,\,\,\,
\frac{1}{\sigma_t}{\sum}_j  E_0^{\ve,\delta\sqrt{\sigma_t}}
|r_t^j|^2 \Big|_{t=\tau/\ve}\to 0\,\,\,\,\mbox{as }\,\,\ve\to0.
$$
Here $\sigma_t\equiv {\sum}_j E_0^\ve |r^j_t|^2$,
and $E_0^{\ve,a} f\equiv E_0^\ve (X^a f)$,
where $X^a$ is the indicator of the event $|f|>a^2$.
Note that (\ref{corf}) and (\ref{5.*}) imply  that
$\sigma_{\tau/\ve} \to{\cal Q}_{\tau,r}(\Psi, \Psi)\not= 0$ as $\ve\to0$.
Hence, it remains to verify the limit relation
$$
{\sum}_j E_0^{\ve,a} |r_{\tau/\ve}^j|^2 \to 0\,\,\,\mbox{as }\,\ve\to0
\,\,\,\,\mbox{ for any }\, a>0.
$$
This condition can be proved by using the technique of \cite{DS}.
\bo

%%%%%%%%%%%%%%%%%%%%%     %%%%%%%%%%%%
\setcounter{equation}{0}
\section{Appendix A: Proof of Lemma 3.6}
%%-------------------------
%We repeat the main steps of the proof for $d=n=1$
%from \ci[Theorem 3.1]{DPST}.
%The difference is not only that
% the dimensions $n,d$ are greater.
%The proof in \ci{DPST} is based on the uniform
% asymptotics of ${\cal G}_t(x)$. 
%We don't analyze this asymptotics in the case $d,n>1$. 
%Instead of it we divide
%${\cal G}_t(x)$ into two summands ${\cal G}^f_t(x)$
%and ${\cal G}^g_t(x)$ and use the asymptotics
%${\cal G}^g_t(x)$ (see Lemma \ref{l5.3}),
%since the contribution of ${\cal G}^f_t(x)$ is negligible
%uniformly on $t$, by (\ref{3.12}).
%%----------------

{\it Step (i)}. Let us study the sum in (\ref{4.9})
over $I_{mN_\ve}\equiv\{x\in\Z^d:
(m_j-1/2)N_\ve\le x_j<(m_j+1/2)N_\ve,\,j=1,\dots d\}$:
$$
\sum\limits_{x\in I_{m N_\ve}}e^{ix\cdot(\theta'-\theta)}=
\prod\limits_{j=1}^d\frac{F(\theta'_j-\theta_j,N_\ve,m_j)}
{e^{i(\theta'_j-\theta_j)}-1},
$$
where
$F(\theta_j,N_\ve,m_j):=
\exp\{i\theta_j N_\ve(m_j+1/2)\}-
\exp\{i\theta_j N_\ve(m_j-1/2)\}$.
Define the function $\al(z)$:
$\ds\al(z)=\frac{iz}{e^{iz}-1}$ if $z\in(-\pi,\pi)\setminus0$
and $\al(0)=1$.
Changing variables in (\ref{4.9}): 
$(\theta,\theta')\to (\theta-z,\theta)$, we obtain
 \beqn\la{3.25}
S^+_{\ve,\tau/{\ve}}&=&
(2\pi)^{-2d}\frac12\sum\limits_{m\in J}
\int\limits_{[-\pi,\pi]^{2d}}e^{-i(l-p)\cdot\theta +il\cdot z}
\prod\limits_{j=1}^d \frac{\al(z_j)F(z_j,N_\ve,m_j)}{iz_j}\nonumber\\
&&\times \hat{\cal G}^g_{\tau/{\ve}}(\theta-z)
\hat {\bf R}_0(\kappa_{r,\ve,m}, \theta)
\hat{\cal G}^{g}_{\tau/{\ve}}(\theta)^*d\theta dz,
\eeqn
where
$\kappa_{r,\ve,m}:=\ve[r/\ve]-\ve m N_\ve$.
Let $C(\theta)$ be defined by (\ref{C(theta)}) 
and $I$ be the identity matrix. Then
$$
\hat{\cal G}^g_{t}(\theta)=g(\theta)\Big(
\cos\Om(\theta)t\, I+\sin\Om(\theta)t\, C(\theta)
\Big),
$$
by (\ref{4.5}) and (\ref{frepecut'}). Let us define
$$
\hat{\cal G}^g_{t,\sigma}(\theta)=g(\theta)\Big(
\cos\om_\sigma(\theta)t\, I+\sin\om_\sigma(\theta)t\, C_\sigma(\theta)
\Big)\quad \mbox{ with } 
C_\sigma(\theta)=\left(\ba{cc}0&\om^{-1}_\sigma(\theta)\\
-\om_\sigma(\theta)&0\ea\right).
$$
Hence,  applying the projections $\Pi_\sigma(\theta)$ 
from Lemma \ref{lc*}, we rewrite the product of matrices 
in the integrand from (\ref{3.25})
as (for $t=\tau/\ve$)
\beqn\la{3.25'}
\hat{\cal G}^g_{t}(\theta-z)
\hat {\bf R}_0(\kappa_{r,\ve,m},\theta)\hat{\cal G}^{g}_{t}(\theta)^*
\!\!&=&\!\!\sum\limits_{\sigma,\sigma'=1}^s\Pi_\sigma(\theta-z)
\hat{\cal G}^g_{t,\sigma}(\theta-z)
\hat {\bf R}_0(\kappa_{r,\ve,m},\theta) 
\hat{\cal G}^g_{t,\sigma'}(\theta)^*\Pi_{\sigma'}(\theta)\nonumber\\
\!\!&=&\!\!
\sum\limits_{\sigma,\sigma'=1}^s\Pi_\sigma(\theta-z)g(\theta-z)
\Big(\sum\limits_{\pm}e^{\pm i\omega_\sigma(\theta-z)t}
\frac{I\mp iC_\sigma(\theta-z)}2\Big) \nonumber\\
&&\!\!\!\!\times
\hat {\bf R}_0(\kappa_{r,\ve,m},\theta) g(\theta)
\Big(\sum\limits_{\pm}e^{\pm i\omega_{\sigma'}(\theta)t}
\frac{I\mp iC^*_{\sigma'}(\theta)}2\Pi_{\sigma'}(\theta)\Big).
\eeqn
{\it Step (ii)}.
Let us consider the one of the terms in (\ref{3.25})
(denote it by $I_\ve^\pm$).
The proof for the remaining terms is similar.
 \beqn\la{3.26}
I^\pm_\ve&:=&(2\pi)^{-2d}\frac18\sum\limits_{m\in J}
\int\limits_{[-\pi,\pi]^{d}}e^{-i(l-p)\cdot\theta}
e^{i \om_{\sigma'}(\theta)\tau/{\ve}}g(\theta)
\Big(\int\limits_{-\pi}^\pi
e^{iz_d l_d} \frac{\al(z_d)F(z_d,N_\ve,m_d)} {iz_d}\ldots\nonumber\\
&&\times\Big(\int\limits_{-\pi}^\pi
e^{iz_2 l_2} \frac{\al(z_2)F(z_2,N_\ve,m_2)}{iz_2}
\Big(\int\limits_{-\pi}^\pi
e^{iz_1 l_1} \frac{\al(z_1)F(z_1,N_\ve,m_1)}
{iz_1}e^{\pm i \om_\sigma(\theta-z)\tau/{\ve}}
\nonumber\\
&&\times
g(\theta-z)\Pi_{\sigma}(\theta-z)\hat{\bf R}_0(\kappa_{r,\ve,m},\theta)
\Pi_{\sigma'}(\theta)dz_1\Big)dz_{2}\Big)\dots dz_d\Big)d\theta.
\eeqn
Let us write
$\nu_1\equiv\nu_1(\theta_1,\theta_2-z_2,\dots)=
\pm[\nabla_1\om_\sigma(\theta_1,\theta_2-z_2,\dots)\tau/(\ve N_\ve)]$,
$\nu_2\equiv\nu_2(\theta_1,\theta_2,\theta_3-z_3,\dots)=
\pm[\nabla_2\om_\sigma(\theta_1,\theta_2,\theta_3-z_3,...)\tau/(\ve N_\ve)]$,
...,
$\nu_d\equiv\nu_d(\theta)=\pm[\nabla_d\om_\sigma(\theta)\tau/(\ve N_\ve)]$.
The first step  in the evaluating the limit value of $I_\ve^\pm$
is the following assertion.
%%%%%%%%%%%%%%%%%%%%%%%%%%%%%%%%%%%%%%%%%%%%%
\begin{pro}\la{ap.pro1} Let condition {\bf I4} hold. Then
 \beqn\la{3.27}
I^\pm_\ve\!\!&=&\!\!(2\pi)^{-2d}\frac18
\int\limits_{[-\pi,\pi]^{d}}e^{-i\theta\cdot(l-p)}
e^{i\om_{\sigma'}(\theta)\tau/{\ve}}g(\theta)
\Big(\sum\limits_{|m_d-\nu_d|\le 2}
\int\limits_{-\pi}^\pi e^{iz_d l_d}
\frac{\al(z_d)F(z_d,N_\ve,m_d)}
{iz_d}\ldots\nonumber\\
&&\!\!\!\!\times\Big(\sum\limits_{|m_2-\nu_2|\le 2} 
\int\limits_{-\pi}^\pi e^{iz_2 l_2}
\frac{\al(z_2)F(z_2,N_\ve,m_2)}
{iz_2}\Big(\sum\limits_{|m_1-\nu_1|\le 2} 
\int\limits_{-\pi}^\pi e^{iz_1 l_1}
\frac{\al(z_1)F(z_1,N_\ve,m_1)}{iz_1}
\nonumber\\
&&\!\!\!\!\times e^{\pm i \om_\sigma(\theta-z)\tau/{\ve}}
\Pi_{\sigma}(\theta-z)\hat{\bf R}_0(\kappa_{r,\ve,m},\theta)
\Pi_{\sigma'}(\theta)g(\theta\!-\!z)
dz_1\Big) dz_2\Big)\dots dz_d \Big)d\theta+o_\tau(1),
\eeqn
where $o_\tau(1)\to 0$ as $\ve\to0$
for any $\tau\not=0$. 
\end{pro}
%%------------------------------------
{\bf Proof}. We generalize the strategy of the proof
of Proposition 3.6 from \ci{DPST},
where this assertion is proved for $d=1$.
To prove the asymptotics (\ref{3.27}), we will show that the series
in (\ref{3.26}) over $\max_j|m_j-\nu_j|\ge 3$
vanishes as $\ve\to0$.
%%%%%%%%%%%%%%%%%%%%%%%%%%%%%%%%%%%%%%%%%%%

Write $J_0=\{n\in\Z^1:|n|\le[c\tau/(\ve N_\ve)]+1\}$ and
$h_j=m_j-\nu_j$. 
Note that in integrand from (\ref{3.26}) the elements of matrix product
have of the form
$$
\Big(\Pi_{\sigma}(\theta-z)\hat{\bf R}_0(\kappa_{r,\ve,m},\theta)
\Pi_{\sigma'}(\theta)\Big)_{\alpha \gamma}
=\!\!\sum\limits^d_{\alpha,\beta,\gamma,\delta=1}
\!\!\!\!\Pi_{\sigma,\alpha\beta}(\theta-z)
\hat{\bf R}_{0,\beta\gamma}(\kappa_{r,\ve,m},\theta)
\Pi_{\sigma',\gamma\delta}(\theta),\,\,\alpha,\gamma=1,\dots,d.
$$
For simplicity of exposition, we omit the sum over
 $\alpha, \beta,\gamma, \delta$ and assume that $d=2$.
Let us denote $\phi_1(\theta)=g(\theta)\Pi_{\sigma,\alpha\beta}(\theta)$,
$\phi_2(\theta)=g(\theta)\Pi_{\sigma',\gamma\delta}(\theta)$, 
and ${\cal R}(h_1,h_2,\theta)=
\hat {\bf R}_{0,\beta\gamma}(\ve[r_1/\ve]-(\nu_1+h_1)\ve N_\ve,
\ve[r_2/\ve]-(\nu_2+h_2)\ve N_\ve,\theta)$. 
Hence, instead of $I_\ve^\pm$ we evaluate the following integral:
\beqn\la{ap1.2}
I'_\ve\!\!&=&\!\!C\sum\limits_{h_2\in J_0-\nu_2}
\sum\limits_{h_1\in J_0-\nu_1} \int\limits_{[-\pi,\pi]^2}
e^{-i\theta\cdot(l-p)}e^{i\om_{\sigma'}(\theta)\tau/{\ve}}\phi_2(\theta)
\Big(
\int\limits_{-\pi}^\pi e^{iz_2(l_2+N_\ve \nu_2)}
\frac{\al(z_2)F(z_2,N_\ve,h_2)}{iz_2}
\nonumber\\
&&\!\!\times\Big(
\int\limits_{-\pi}^\pi e^{iz_1(l_1+ N_\ve \nu_1)}
\frac{\al(z_1)F(z_1,N_\ve,h_1)}{iz_1}
 e^{\pm i \om_\sigma(\theta-z)\tau/{\ve}}
\phi_1(\theta\!-\!z){\cal R}(h_1,h_2,\theta)dz_1\Big)
dz_2\Big)d\theta.\nonumber
\eeqn
Here we use the fact that
$F(z_j,N_\ve,m_j)=e^{iz_j N_\ve \nu_j}F(z_j,N_\ve,h_j)$.
Decompose the series over $h_1$ and $h_2$ in $I'_\ve$ into the sums:
over $h_j\le -3$, over $|h_j|\le2$ and $h_j\ge 3$, $j=1,2$.
Therefore,
\be\la{6.4'}
I'_\ve=\sum\limits_{i,j=1}^3 I_\ve^{(i,j)},
\ee
where
$$
I_\ve^{(1,1)}=\!\!\!\!\sum\limits_{\scriptsize{
\ba{cc} h_1\in J_0-\nu_1\\h_1\le -3\ea}}\!\!\!\!
\sum\limits_{\scriptsize{
\ba{cc} h_2\in J_0-\nu_2\\h_2\le -3\ea}}\!\!\!\!\!\!\!\dots,\,\,\,
I_\ve^{(1,2)}=\!\!\!\!\sum\limits_{\scriptsize{
\ba{cc} h_1\in J_0-\nu_1\\h_1\le -3\ea}}\!\!\!\!
\sum\limits_{\scriptsize{
\ba{cc} h_2\in J_0-\nu_2\\|h_2|\le 2\ea}}\!\!\!\!\!\!\!\dots,\,\,\,
I_\ve^{(1,3)}=\!\!\!\!\sum\limits_{\scriptsize{
\ba{cc} h_1\in J_0-\nu_1\\h_1\le -3\ea}}\!\!\!\!
\sum\limits_{\scriptsize{
\ba{cc} h_2\in J_0-\nu_2\\h_2\ge 3\ea}}\!\!\!\!\!\!\!\dots,
$$
and so on. We want to prove that the series in $I'_\ve$
over $\max_j|h_j|\ge3$ 
vanishes as $\ve\to0$, i.e., $I_\ve^{(i,j)}$ vanish as $\ve\to0$
if $i=1,3$ or $j=1,3$.
We prove this fact only for $I_\ve^{(1,1)}$.
For remaining integrals  the proof is similar.
%%%%%%%%%%%%%%%%%%%%%%%%%%%%%%%%%%%%%%%%%%%
Let us write 
\beqn\nonumber
h^j_{min}=\min\{h_j\in\Z^1:h_j\le-3\,\,\,\mbox{and }\,h_j\in J_0-\nu_j\}.
\eeqn
%Therefore, $h^j_{min}$ is such negative integer number that
%$h^j_{min}\in[-\tau/(\ve N_\ve)-1/2-\nu_j,-\tau/(\ve N_\ve)+1/2-\nu_j)$, $j=1,2$.
Hence,
\beqn
I_\ve^{(1,1)}\!\!\!&=&\!\!\!C\!\int\limits_{[-\pi,\pi]^2}
\!\!\!\!e^{-i\theta\cdot(l-p)}e^{i\om_{\sigma'}(\theta)\tau/{\ve}}
\phi_2(\theta)
\Big(\sum\limits_{h^2_{min}\le h_2\le -3}
\int\limits_{-\pi}^\pi e^{iz_2(l_2+N_\ve\nu_2)}
\frac{\al(z_2)F(z_2,N_\ve,h_2)}{iz_2} 
I_1\, dz_2\Big)d\theta,\nonumber
\eeqn
where
\beqn\la{ap1.3}
I_1\!\!\!&\equiv&\!\!\! I_1(\theta,z_2,h_2)\nonumber\\
&=&\!\!\!\!\!\sum\limits_{h^1_{min}\le h_1\le -3}
\int\limits_{-\pi}^\pi e^{iz_1(l_1+ N_\ve \nu_1)}
\frac{\al(z_1)F(z_1,N_\ve,h_1)}{iz_1} 
 e^{\pm i \om_\sigma(\theta\!-\!z)\tau/{\ve}}
\phi_1(\theta\!-\!z)\,dz_1{\cal R}(h_1,h_2,\theta).
\eeqn
To rewrite the sums over $h_1$ and $h_2$
we use the following "discrete integration-by-parts formula"
(see \cite[p.594]{DPST})
\beqn\la{ap1.4}
&&\sum\limits_{h_{min}\le h\le -3}
\Big[e^{i(h+1/2)N_\ve z}-e^{i(h-1/2)N_\ve z}\Big]
f(h)= \Big[e^{-i5/2N_\ve z}-e^{i(h_{min}-1/2)N_\ve z}\Big]f(-3)
\nonumber\\
&&~~~~~~~~~~~~~~+\sum\limits_{h_{min}\le h\le -4} 
\!\!\!\!\Big[e^{i(h+1/2)N_\ve z}-e^{i(h_{min}-1/2)N_\ve z}\Big]
\Big(f(h)-f(h+1)\Big).
\eeqn
%%%%%%%%%%%%%%%%%%%%%%%%%%%%%%%%%%%%%%%
%Indeed,
%\beqn\nonumber
%S_1&=&\sum\limits_{h^1_{min}\le h_1\le -3} 
%[e^{i(h_1+1/2)N_\ve z_1}-e^{i(h^1_{min}-1/2)N_\ve z_1}+
%e^{i(h^1_{min}-1/2)N_\ve z_1}-e^{i(h_1-1/2)N_\ve z_1}]
%{\cal R}(h_1,h_2,\theta)\\
%&=&[e^{-i5/2N_\ve z_1}-e^{i(h^1_{min}-1/2)N_\ve z_1}]{\cal R}(-3,h_2,\theta)\\
%&&+\sum\limits_{h^1_{min}\le h_1\le -4} 
%\Big[e^{i(h_1+1/2)N_\ve z_1}-e^{i(h^1_{min}-1/2)N_\ve z_1}\Big]
%{\cal R}(h_1,h_2,\theta)\nonumber\\
%&&-\sum\limits_{h^1_{min}\le h_1\le -3}
%\Big[e^{i(h_1-1/2)N_\ve z_1}-e^{i(h^1_{min}-1/2)N_\ve z_1}\Big] 
%{\cal R}(h_1,h_2,\theta).\nonumber
%\eeqn
%Replacing $h_1\to h_1-1$ in the last sum we obtain (\ref{ap1.4}).
%%%%%%%%%%%%%%%%%%%%%%%%%%%%%%%%%%%%%%%%%%%%%%
Let us apply (\ref{ap1.4}) to the sum over $h_2$:
\beqn
&&\sum\limits_{h^2_{min}\le h_2\le -3}
F(z_2,N_\ve,h_2) I_1(\theta,z_2,h_2)=\Big[
e^{-i5/2N_\ve z_2}-e^{i(h^2_{min}-1/2)N_\ve z_2}\Big]
I_1(\theta,z_2,-3)
\nonumber\\
&&+\sum\limits_{h^2_{min}\le h_2\le -4} 
\!\!\!\!\Big[e^{i(h_2+1/2)N_\ve z}-e^{i(h^2_{min}-1/2)N_\ve z_2}\Big]
\Big(I_1(\theta,z_2,h_2)-I_1(\theta,z_2,h_2+1)\Big).
\eeqn\nonumber
Hence,
\beqn\la{ap1.4.1}
|I_\ve^{(1,1)}|\le C_1
\sup_{\theta\in[-\pi,\pi]^2}\sup_{z_2\in[-\pi,\pi]}
\Big(|I_1(\theta,z_2,-3)|+\sum\limits_{h^2_{min}\le h_2\le -4} 
\!\!\!\Big|I_1(\theta,z_2,h_2)-I_1(\theta,z_2,h_2+1)\Big|\Big).
\eeqn
Applying the formula (\ref{ap1.4}) to the sum over $h_1$,
we rewrite $I_1$ in the form 
\beqn\la{ap1.5}
I_1&=&C(\theta,z_2,-5/2,h^1_{min}-1/2){\cal R}(-3,h_2,\theta)
\nonumber\\
&&+\sum\limits_{h^1_{min}\le h_1\le -4} C(\theta,z_2,h_1+1/2,h^1_{min}-1/2)
\Big({\cal R}(h_1,h_2,\theta)
-{\cal R}(h_1+1,h_2,\theta)\Big),\,\,\,\,\,
\eeqn
where, by definition, the function $C(\theta,z_2,m,m')$ is equal to
\beqn\la{ap1.6}
C(\theta,z_2,m,m')=
\int\limits_{-\pi}^\pi e^{iz_1(l_1+\nu_1 N_\ve)
\pm i \om_\sigma(\theta-z)\tau/{\ve}}
 \frac{\al(z_1)}{iz_1}(e^{imN_\ve z_1}-e^{im'N_\ve z_1})
\phi_1(\theta-z)\,dz_1.
\eeqn
Substituting (\ref{ap1.5}) in (\ref{ap1.4.1}) we obtain
\beqn\la{ap1.6.1}
|I_\ve^{(1,1)}|&\le& C_1
\sup_{\theta\in[-\pi,\pi]^2}\sup_{z_2\in[-\pi,\pi]}
\Big[\sup_{h^1_{min}\le h_1\le -3}
|C(\theta,z_2,h_1+1/2,h^1_{min}-1/2)| \nonumber\\
&&\times\Big(\sum\limits_{h^1_{min}\le h_1\le -4} 
\Big|{\cal R}(h_1,-3,\theta)-{\cal R}(h_1+1,-3,\theta)\Big|
\nonumber\\
&&+\sum\limits_{h^2_{min}\le h_2\le -4} 
\Big|{\cal R}(-3,h_2,\theta)
-{\cal R}(-3,h_2+1,\theta)\Big| \nonumber\\
&&+\!\!
\sum\limits_{h^1_{min}\le h_1\le -4} \sum\limits_{h^2_{min}\le h_2\le -4} 
|{\cal R}(h_1,h_2,\theta)-
{\cal R}(h_1+1,h_2,\theta)\nonumber\\
&&~~~-{\cal R}(h_1,h_2\!+\!1,\theta)
+{\cal R}(h_1\!+\!1,h_2\!+\!1,\theta)|\Big)\Big].\nonumber
\eeqn
By condition {\bf I4}, all sums in (\ref{ap1.6.1}) are bounded
uniformly on $\theta$, since they do not exceed the variations 
of $\hat{\bf R}_0(\cdot,\theta)$ on the set
$[r_1-4-c\tau,r_1+4+c\tau]\times[r_2-4-c\tau,r_2+4+c\tau]$. 
Hence,
\beqn\la{ap1.6.2}
|I_\ve^{(1,1)}|\le C_2
\sup_{\theta\in[-\pi,\pi]^2}\sup_{z_2\in[-\pi,\pi]}
\sup_{h^1_{min}\le h_1\le -3}
|C(\theta,z_2,h_1+1/2,h^1_{min}-1/2)|.
\eeqn
%%%%%%%%%%%%%%%%%%%%%%%%%%%%%%%%%%
\begin{lemma} \la{l10.1}
(see Lemma 3.7 from \cite{DPST})
Let condition {\bf I1} (ii) hold. Then
\beqn\la{ap1.7}
e^{\mp i \om_\sigma(\theta_1,\theta_2-z_2)\tau/{\ve}}C(\theta,z_2,m,m')
\to \pi(\sign m-\sign m')\phi_1(\theta_1,\theta_2-z_2),\,\,\,\,\ve\to0,
\eeqn
uniformly in $\theta\in[-\pi,\pi]^2$, $z_2\in [-\pi,\pi]$ and $|m|,|m'|>2$.
\end{lemma}
%%%%%%%%%%%%%%%%%%%%%%%%%%%%%%%%%

Since $h_1+1/2\le-2$, $h^1_{min}-1/2\le-2$, 
the integral $I_\ve^{(1,1)}$ vanishes as $\ve\to0$ by Lemma~\ref{l10.1} and 
(\ref{ap1.6.2}).
Similarly, the remaining integrals in (\ref{6.4'}) with $i=1,3$ or $j=1,3$,
vanish as $\ve\to0$, i.e., the series in $I'_\ve$ over 
$\max_j|h_j|\ge3$ vanish as $\ve\to0$.
Proposition \ref{ap.pro1} is proved.\bo
\medskip\\
%%%%%%%%%%%%%%%%%%%%%%%%%%%%%%%%%%%%%%%%%%%%%%%%%%%
{\it Step (iii)}
 The next step in the proof is to prove the following asymptotics 
for the RHS of (\ref{3.27}) as $\ve\to0$:
 \beqn\la{3.29}
I^\pm_\ve\!&=&\!\frac{(2\pi)^{-2d}}8
\int\limits_{[-\pi,\pi]^{d}}
e^{-i\theta\cdot(l-p)+i \om_{\sigma'}(\theta)\tau/{\ve}}
g(\theta)\Pi_{\sigma}(\theta)
\hat{\bf R}_0(r\mp\nabla\om_\sigma(\theta)\tau,\theta)
\Pi_{\sigma'}(\theta)\nonumber\\
&&\!\!\!\!\Big(\int\limits_{-\pi}^\pi
e^{iz_d (l_d+\nu_d N_\ve)} 
\al(z_d)\frac{e^{i5/2 N_\ve z_d}\!-\!
e^{-i5/2 N_\ve z_d}}{iz_d}\Big(\ldots
\Big(\int\limits_{-\pi}^\pi e^{iz_2(l_2+\nu_2 N_\ve)} 
\al(z_2)\frac{e^{i5/2 N_\ve z_2}-
e^{-i5/2 N_\ve z_2}}{iz_2}\nonumber\\
&&\!\!\!\!\Big(\int\limits_{-\pi}^\pi
e^{iz_1 (l_1+\nu_1 N_\ve)\pm i \om_\sigma(\theta-z)\tau/{\ve}} 
\al(z_1)\frac{e^{i5/2 N_\ve z_1}-
e^{-i5/2 N_\ve z_1}}{iz_1}
\,dz_1\Big) dz_2\Big)\dots\Big) dz_d\Big)d\theta\!+\!o_\tau(1),
\eeqn
where $o_\tau(1)\to 0$ as $\ve\to0$
for any $\tau\not=0$. 
%%%%%%%%%%%%%%%%%
 Formula (\ref{3.29}) was proved
in \ci[Lemma 3.8]{DPST} for the case when $d=1$.
This formula is based on
the formula
$\sum_{|m_j-\nu_j|\le2}F(z_j,N_\ve,m_j)=
e^{i\nu_j N_\ve z_j}(e^{i5/2 N_\ve z_j}-e^{-i5/2 N_\ve z_j})$
and  the following inequality
$$
|\hat{\bf R}_0(\ve[r/\ve]-\nu\ve N_\ve-h\ve N_\ve,\theta)
-\hat{\bf R}_0(r\mp \nabla\omega_\sigma(\theta)\tau,\theta)|\le C(\ve+N_\ve).
$$
This inequality follows because  $|h|\le 2$ and 
$\hat{\bf R}_0(r,\theta)$ satisfies condition {\bf I4}.
The proof of \ci[Lemma 3.8]{DPST} admits extension to the case
when $d>1$.
\medskip

%%%%%%%%%%%%%%%%%%%%%%%%%%%%%%%%%%%%%%%%%%%
{\it Step (iv)}
Let us apply Lemma \ref{l10.1}
to the inner integrals in the RHS of (\ref{3.29})
over $z_1,\dots,z_d$ and obtain
 \beqn\la{3.30}
I^\pm_\ve=\frac{(2\pi)^{-d}}8
\int\limits_{\T^{d}}e^{-i\theta\cdot(l-p)}
e^{i \big(\om_{\sigma'}(\theta)\pm \om_\sigma(\theta)\big)\tau/{\ve}}
g(\theta)\Pi_{\sigma}(\theta)
\hat{\bf R}_0(r\mp\nabla\om_\sigma(\theta)\tau,\theta)\Pi_{\sigma'}(\theta)
\,d\theta+o(1),
\eeqn
 as $\ve\to0$. 
%%%%%%%%%%%%%%%%%%%%%%%%%%%%%%%%%%%%%%%%%%%
Note that $\hat{\bf R}^{ij}_0(r,\cdot)\in C(\T^d)$,
$i,j=0,1$,  by condition (\ref{3.2'}). 
Moreover, the identities 
$\om_\sigma(\theta)+\om_{\sigma'}(\theta)\equiv{\rm const}_+$ or  
$\om_\sigma(\theta)-\om_{\sigma'}(\theta)\equiv{\rm const}_-$  
 with the ${\rm const}_\pm\ne 0$ are 
impossible by condition {\bf E5}. 
Furthermore, the oscillatory integrals 
with $\om_\sigma(\theta)\pm\om_{\sigma'}(\theta)\not\equiv{\rm const}_\pm$
vanish as $\ve\to0$ by %condition {\bf I1} and 
the Lebesgue--Riemann theorem.
Hence, only the integrals with $\om_\sigma(\theta)-\om_{\sigma'}(\theta)
\equiv 0$  contribute to the integral  (\ref{3.30}),  
since  $\om_\sigma(\theta)+\om_{\sigma'}(\theta)\equiv 0$ would imply  
$\om_\sigma(\theta)\equiv\om_{\sigma'}(\theta)\equiv 0$, which 
is impossible by  {\bf E4}. 
Thus, for $\sigma\not=\sigma'$, $I_\ve^\pm=o(1)$ as $\ve\to0$.
For $\sigma=\sigma'$,  $I_\ve^+=o(1)$ and
$$
I^-_\ve=\frac{(2\pi)^{-d}}8
\int\limits_{\T^{d}}e^{-i\theta\cdot(l-p)}
g(\theta)\Pi_{\sigma}(\theta)
\hat{\bf R}_0(r+\nabla\om_\sigma(\theta)\tau,\theta)\Pi_{\sigma'}(\theta)
\,d\theta+o(1)\,\,\,\mbox{as }\,\ve\to0. 
$$
This completes the proof of Lemma \ref{Qt1}.
\bo
%%%%%%%%%%%%%%%%%%%%%%%%%%%%%%%%%%%%%%%%%%%%%%%%   
%%%%%%%%%%%%%%%%%%%%%     %%%%%%%%%%%%
\setcounter{equation}{0}
\section{Appendix B: Proof of Lemma 3.7}
%%-------------------------
We first apply (\ref{d1''})  and obtain
$$
S^-_{\ve,\tau/\ve}
=\frac12\sum\limits_{m\in J}\sum\limits_{x\in I_{m N_\ve}}
{\cal G}^g_{\tau/\ve}(l+x)\sum\limits_{y\in \Z^d}
 {\bf R}_0(\ve[r/\ve]-\ve m N_\ve, y-x)\sign([r_1/\ve]-y_1)
{\cal G}^g_{\tau/\ve}(p+y)^T.
$$
Let us write $\kappa_{r,\ve,m}=\ve[r/\ve]-\ve m N_\ve$. 
The Parseval equality yields
\beqn
&&\sum\limits_{y\in \Z^d}\sign([r_1/\ve]-y_1)
{\bf R}_0(\kappa_{r,\ve,m}, y-x){\cal G}^g_{\tau/\ve}(p+y)^T\nonumber\\
&=&(2\pi)^{-d}\int\limits_{\T^{d}}
F_{y\to\theta'}\Big[\sign([r_1/\ve]-y_1)
 {\bf R}_0(\kappa_{r,\ve,m}, y-x)\Big]
\overline{F_{y\to\theta'}[{\cal G}^g_{\tau/\ve}(p+y)^T]}\,d\theta'.
\eeqn
Note that
$$
F_{y\to\theta'}[\sign([r_1/\ve]-y_1)]=-i\,(2\pi)^{d-1}
\delta(\bar \theta')\PV\left(\frac1{\tg(\theta'_1/2)}\right)
e^{i[r_1/\ve]\theta'_1},
$$
where $\theta'=(\theta'_1,\bar\theta')$, and
$\PV$ stands for the Cauchy principal part.
Hence,
 \beqn
S^-_{\ve,\tau/\ve}&=&
-\frac{i}{2}(2\pi)^{-d-1}\sum\limits_{m\in J}
\sum\limits_{x\in I_{m N_\ve}}{\cal G}^g_{\tau/\ve}(l+x)
\int\limits_{\T^{d}}\left(\PV\int\limits_{\T^1}
\frac{e^{i[r_1/\ve](\theta'_1-z)+ix_1z}}{\tg((\theta'_1-z)/2)}
\hat {\bf R}_0(\kappa_{r,\ve,m},z,\bar\theta')dz\right)\nonumber\\
&&
\times e^{i\bar x\cdot\bar\theta'}
\hat{\cal G}^g_{\tau/\ve}(\theta')^*e^{ip\cdot\theta'}\,d\theta'\nonumber\\
&=&-\frac{i}2(2\pi)^{-2d-1}\sum\limits_{m\in J}
\sum\limits_{x\in I_{m N_\ve}}\int\limits_{\T^{2d}}
e^{-i(l+x)\cdot\theta} \hat{\cal G}^g_{\tau/\ve}(\theta)
\nonumber\\
&&\times \left(\PV\int\limits_{\T^1}
\frac{e^{i[r_1/\ve](\theta'_1-z)+ix_1z}}{\tg((\theta'_1-z)/2)}
\hat {\bf R}_0(\kappa_{r,\ve,m},z,\bar\theta')\,dz\right)
e^{i\bar x\cdot\bar\theta'}\hat{\cal G}^g_{\tau/\ve}(\theta')^*
e^{ip\cdot\theta'}\,d\theta'd\theta.
\eeqn 
We change variables: $\theta'_1\to\varphi=\theta'_1-z$,
and then denote $z=\theta'_1$. Therefore,
\beqn\nonumber
S^-_{\ve,\tau/\ve}=\frac{(2\pi)^{-2d}}{4\pi i}
\sum\limits_{m\in J}\sum\limits_{x\in I_{m N_\ve}}
\int\limits_{\T^{2d}}
e^{-i(l\cdot\theta-p\cdot\theta')}
e^{i x\cdot(\theta'-\theta)}
\hat{\cal G}^g_{\tau/\ve}(\theta)
\hat {\bf R}_0(\kappa_{r,\ve,m},\theta')
I_{\ve}(\theta')\,d\theta'd\theta,
\eeqn
where
$$
I_{\ve}(\theta'):=\PV\int\limits_{\T^1}
\frac{e^{i([r_1/\ve]+p_1)\varphi}}{\tg(\varphi/2)}
\hat{\cal G}_{\tau/\ve}(\theta'_1+\varphi,\bar\theta')^*
g(\theta'_1+\varphi,\bar\theta')\,d\varphi,\,\,\,
\theta'=(\theta'_1,\bar\theta').
$$
By formula (\ref{4.5}), the matrix  $\hat{\cal G}_{t}(\theta)$ has the form
$$
\hat{\cal G}_{t}(\theta)=
\cos\Omega(\theta)t \,I+\sin\Omega(\theta)t\, C(\theta),
$$
where  $I$ is the identity matrix and
 $C(\theta)$ is introduced in  (\ref{C(theta)}).
By Lemma \ref{lc*} (iv),
\beqn\la{4.15}
\hat{\cal G}^*_{t}(\theta)&=&\sum\limits_{\sigma=1}^s
(\cos\omega_\sigma(\theta)t +\sin\omega_\sigma(\theta)t\, C_\sigma^*(\theta))
\Pi_\sigma(\theta)
\nonumber\\
&=&\sum\limits_{\sigma=1}^s
e^{\pm i\omega_\sigma(\theta)\tau/\ve}
\frac{I\mp i C^*_\sigma(\theta)}{2}\Pi_\sigma(\theta),
\quad \theta\in\T^d\setminus {\cal C}_*,
\eeqn
since $\cos\omega_\sigma(\theta)t
=(e^{i\omega_\sigma t}+e^{-i\omega_\sigma t})/2$
and $\sin\omega_\sigma(\theta)t=
(e^{i\omega_\sigma t}-e^{-i\omega_\sigma t})/(2i)$.
Applying the partition of unity (\ref{part}), (\ref{fge}),
and formula (\ref{4.15})
we rewrite $I_\ve(\theta)$ in the form 
\beqn\nonumber
I_\ve(\theta)=\sum\limits_{k,\pm}
\sum\limits_{\sigma=1}^s
\PV\int\limits_{\T^1} g_k(\theta_1\!+\!\varphi,\bar\theta)
\frac{e^{i([r_1/\ve]+p_1)\varphi}
e^{\pm i\omega_\sigma(\theta_1+\varphi,\bar \theta)\tau/\ve}}{\tg(\varphi/2)}
\frac{I\mp i C^*_\sigma(\theta_1\!+\!\varphi,\bar\theta)}{2}
\Pi_\sigma(\theta_1\!+\!\varphi,\bar\theta)\,d\varphi.
\eeqn
%%%%%%%%%%%%%%%%%%%%%%%%%%%%%%
\begin{lemma}\label{appendixA}
(i) $\sup\limits_{\theta\in\T^d,r_1\in\R}\sup_{\ve>0}
|I_{\ve}(\theta)|<\infty$.\\
(ii) Let $\nabla_1\omega_\sigma(\theta)\not=\pm r_1$,
for $\theta\in\supp g_k$ and for fixed $r_1\in\R$.
Then
\be\label{5.6}
I_\ve(\theta)-2\pi i\sum\limits_{k,\pm}
\sum\limits_{\sigma=1}^s g_k(\theta)
e^{\pm i\omega_\sigma(\theta)\tau/\ve}
f^\pm_{r_1}(\theta)\Pi_\sigma(\theta)\to0
\quad\mbox{as }\,\, \ve\to+0,
\ee
 for fixed $\tau>0$ and $r_1\in\R$.
Here $f^\pm_{r_1}(\theta)$ is a matrix-valued function of the form
$$
f^\pm_{r_1}(\theta)=\sign(r_1\pm\nabla_1\omega_\sigma(\theta)\tau)
(I\mp i C^*_\sigma(\theta))/2.
$$ 
\end{lemma}
%%-------------------------

 Lemma \ref{appendixA} can be proved by using the 
technique of  \cite[Lemma 8.3]{DKM}
or of  \cite[Proposition A.4 (i), (ii)]{BPT}. The proof is
based on the following well-known assertion
$$
\lim_{\lambda\to+\infty}
\left(\PV\int\limits_{-\pi}^{\pi}\frac{e^{i\lambda f(z)}\chi(z)}{z}\,dz-
\pi ie^{i\lambda f(0)}\chi(0)\sign f'(0)\right)=0,
$$
where $\chi\in C^1$, $f\in C^2$ and $f'(0)\not=0$.
%%----------------------------
 Lemma \ref{appendixA} gives
\beqn\la{5.60}
S^-_{\ve,\tau/\ve}&=&(2\pi)^{-2d}\frac12\sum\limits_{\sigma=1}^s
\sum\limits_{k,\pm}\sum\limits_{m\in J}\sum\limits_{x\in I_{m N_\ve}}
\int\limits_{\T^{2d}} g_k(\theta)
e^{-i(l\cdot\theta-p\cdot\theta')}
e^{i x\cdot(\theta'-\theta)}
\hat{\cal G}^g_{\tau/\ve}(\theta)\nonumber\\
&&\times\hat {\bf R}_0(\kappa_{r,\ve,m},\theta')
e^{\pm i\omega_\sigma(\theta')\tau/\ve}f^\pm_{r_1}(\theta')
\Pi_\sigma(\theta')\,d\theta'd\theta.
\eeqn
Comparing (\ref{4.9}) and (\ref{5.60}) we see that
 the problem of evaluating the limit value of (\ref{5.60})
is solved by the similar way as in Lemma \ref{Qt1}. \bo
%%------------------------------------------------

%%%%%%%%%%%%%%%%%%%%%     %%%%%%%%%%%%
\setcounter{equation}{0}
\section{Appendix C: Proof of Lemma 3.8}
%%-------------------------

By (\ref{Qta}) we  write
$$
S^0_{\ve,\tau/\ve}=
\sum\limits_{m\in J}\sum\limits_{x\in I_{mN_\ve}}\sum\limits_{y\in\Z^d}
{\cal G}^g_{\tau/\ve}(l+\!x) R^0(\kappa_{r,\ve,m},
[r/\ve]\!-\!x,[r/\ve]\!-\!y){\cal G}^{g}_{\tau/\ve}(p+y)^T,
$$ 
where $\kappa_{r,\ve,m}=\ve[r/\ve]-\ve m N_\ve$. 
Change variables $y\to z=y-x$ and denote the sum over $m$ and $x$ by
$\Phi_\ve(z)$,
\beqn\label{B.2}
\Phi_\ve(z)&\equiv& \Phi_\ve(z,\tau,r,l,p)\nonumber\\
&=&
\sum\limits_{m\in J}\sum\limits_{x\in I_{mN_\ve}}
{\cal G}^g_{\tau/\ve}(l+\!x) R^0(\kappa_{r,\ve,m},
[r/\ve]\!-\!x,[r/\ve]\!-\!x-z){\cal G}^{g}_{\tau/\ve}(p+x+z)^T.
\eeqn
Therefore,
\be\label{B.1}
S^0_{\ve,\tau/\ve}=\sum\limits_{z\in\Z^d}\Phi_\ve(z).
\ee
The estimate (\ref{3.2}) and definition (\ref{d1'''}) imply the same estimate 
for $R^0$:
\be\la{B.3}
|R^0(r,x,y)|\le C(1+|x-y|)^{-\gamma}.
\ee
Next, the Cauchy--Schwartz inequality yields
\beqn\la{B.4}
\sum\limits_{m\in J}\sum\limits_{x\in I_{mN_\ve}}
|{\cal G}^g_{\tau/\ve}(l+\!x)| |{\cal G}^{g}_{\tau/\ve}(p+x+z)^T|&\le&
\sum\limits_{x\in\Z^d}
|{\cal G}^g_{\tau/\ve}(l+\!x)| |{\cal G}^{g}_{\tau/\ve}(p+x+z)^T|\nonumber\\
&\le& \Vert {\cal G}^g_{\tau/\ve}\Vert^2_{\ell^2}
\le C (1+\Vert \hat V^{-1}\Vert^2_{L^2(\T^d)}).
\eeqn
Hence, condition {\bf E6} and estimate (\ref{B.3}) imply that
$
|\Phi_\ve(z)|\le C(1+|z|)^{-\gamma}.
$
Since $\gamma>d$,
\be\label{B.5}
\sum\limits_{z\in\Z^d}|\Phi_\ve(z)|\le C<\infty,
\ee
and the series in (\ref{B.1}) converges uniformly in $\ve$
(and also in $\tau,r,l,p$). Therefore, it suffices to prove that
\be\la{B.6}
\lim_{\ve\to0}\Phi_\ve(z)=0\quad\mbox{for each }\, z\in\Z^d.
\ee
Let us consider the series  in (\ref{B.2}). At first, note that
by definitions {\bf I0} and (\ref{d1'''}), 
the function $R^0(r,x,y)$ depends on $\bar x-\bar y$, i.e.,
has the form   
$R^0(r,x,y)={\bf R}^0(r,x_1,y_1,\bar x-\bar y)$, and 
${\bf R}^0(r,x_1,y_1,\bar z)=0$ for $y_1<0$. Hence,
$$
{\bf R}^0(r,[r_1/\ve] -x_1,[r_1/\ve]-x_1-z_1,\bar z)=0
\quad \mbox{for }\, x_1\ge [r_1/\ve]-z_1.
$$
Further, from condition (\ref{3.14}) it follows that  $\forall \delta>0$
$\exists K_\delta>0$ such that for any $y_1>K_\delta$
$|{\bf R}^0(r,y_1,y_1-z_1,\bar z)|<\delta$.
Hence, $\forall \delta>0$ $\exists M_\delta=\max(K_\delta,z_1)>0$
such that
\beqn
&&\Big|\sum\limits_{m\in J}\sum\limits_{\scriptsize{
\ba{cc}x\in I_{mN_\ve}\\
x_1<[r_1/\ve]-M_\delta\ea}}\!\!\!\!\!\!\!
{\cal G}^g_{\tau/\ve}(l+\!x) {\bf R}^0(\kappa_{r,\ve,m},
[r_1/\ve]\!-\!x_1,[r_1/\ve]\!-\!x_1-z_1,\bar z)
{\cal G}^{g}_{\tau/\ve}(p+x+z)^T\Big|
\nonumber\\
&\le&\delta \sum\limits_{m\in J}\sum\limits_{x\in I_{mN_\ve}}
\Big|{\cal G}^g_{\tau/\ve}(l+\!x){\cal G}^{g}_{\tau/\ve}(p+x+z)^T\Big|
\le C\delta,
\nonumber
\eeqn 
by estimate (\ref{B.4}).
It remains to prove that for fixed $M_\delta>0$,
\be\la{B.10}
\sum\limits_{m\in J}\sum\limits_{x\in A_m}
{\cal G}^g_{\tau/\ve}(l+\!x) {\bf R}^0(\kappa_{r,\ve,m},
[r_1/\ve]\!-\!x_1,[r_1/\ve]\!-\!x_1-z_1,\bar z)
{\cal G}^{g}_{\tau/\ve}(p+x+z)^T \to0,\,\,\,\ve\to0,
\ee
where $A_m=\{x=(x_1,\bar x):\,
x_1\in([r_1/\ve]-M_\delta,[r_1/\ve]-z_1)\cap I_{m_1N_\ve};
\bar x\in I_{\bar mN_\ve}\}$. For enough small an $\ve>0$, there is a
%%$m^\ve_1=(-\frac12+[\frac{r_1}{\ve}] \frac1{N_\ve}-\frac{z_1}{N_\ve},
%%\frac12+[\frac{r_1}{\ve}] \frac{1}{N_\ve}-\frac{M_\delta}{N_\ve})\in\Z$ 
$m^\ve_1\in\Z$ such that
$([r_1/\ve]-M_\delta,[r_1/\ve]-z_1)\subset I_{m^\ve_1N_\ve}$.
Hence,
$|\ve[r_1/\ve]-\ve m_1^\ve N_\ve|\le C(\ve N_\ve+\ve)$.
Therefore, by condition {\bf I4}, we have
$$
|R^0(\ve[r_1/\ve]-\ve m_1^\ve N_\ve,\dots)-R^0(0,\dots)|\le C\ve^{1-\beta}
\quad\mbox{with some } \,\beta\in(0,1).
$$
Hence, by the estimate (\ref{B.4}), we can replace 
$R^0(\ve[r_1/\ve]-\ve m_1^\ve N_\ve,\dots)$ into $R^0(0,\dots)$
in the series (\ref{B.10}).

Further, let us change $x_1\to [r_1/\ve]-x_1$ in (\ref{B.10}).
Therefore, $x_1$ runs the finite number of points,
$x_1\in(z_1, M_\delta)$. 
To derive (\ref{B.6}) it suffices to prove that for every fixed $z\in\Z^d$,
$x_1\in\Z$, $r\in\R^d$, and $\tau\not=0$,
\be\label{B.11}
\sum\limits_{\bar m\in \bar J}
{\bf R}^{0}(0,\bar r-\ve\bar mN_\ve,x_1,x_1-z_1,\bar z)_{\beta\gamma}
\sum\limits_{\bar z\in I_{\bar m N_\ve}}
{\cal G}^{g}_{\tau/\ve}(l+x_r)_{\alpha\beta} 
{\cal G}^{g}_{\tau/\ve}(p+x_r+z)^T_{\gamma\delta}\to0
\ee
as $\ve\to0$, where $\alpha,\beta,\gamma,\delta=1,\dots,d$,
$x_r:=([r_1/\ve]-x_1,\bar x)$. 
For simplicity of exposition, we omit indices $i,j,k,l$ in (\ref{B.11})
and assume that $d=1$.
In this case, to prove (\ref{B.11}) it suffices to show that
\beqn\nonumber
{\cal G}^{g}_{\tau/\ve}(l_1+[r_1/\ve]-x_1) 
{\cal G}^{g}_{\tau/\ve}(p+[r_1/\ve]-x_1+z)^T\to0\,\,\, \mbox{as }\,\ve\to0.
\eeqn
Indeed, applying the Fourier transform, we have
\be\la{B.12}
{\cal G}^{g}_{\tau/\ve}(l_1+[r_1/\ve]-x_1)=(2\pi)^{-1}
\int_{{\bf T}^1} e^{-i(l_1+[r_1/\ve]-x_1)\theta_1}\hat {\cal G}_{\tau/\ve}(\theta_1)
g(\theta_1)\,d\theta_1.
\ee
Using the decomposition (\ref{frepecut'}) 
we rewrite (\ref{B.12}) as
\be\la{B.13}
{\cal G}^{g}_{\tau/\ve}(l_1+[r_1/\ve]-x_1)=(2\pi)^{-1}\sum_{k,\pm,\sigma}
\int_{{\bf T}^1} g_k(\theta_1)a_\pm(\theta_1)g(\theta_1)
e^{-i(l_1+[r_1/\ve]-x_1)\theta_1\pm i\omega_\sigma(\theta_1)\tau/\ve}
\,d\theta_1.
\ee
The eigenvalues $\om_\sigma(\theta)$
and the matrices $a^\pm_\sigma(\theta)$ are real-analytic functions inside
the $\supp g_k$ for every $k$. 
Moreover, conditions {\bf E4} and {\bf E6} imply that for fixed $r_1\in\R^1$
and $\tau\not=0$,
mes$\{\theta_1\in\T^1:\,\nabla_1\omega_\sigma(\theta_1)=\pm r_1/\tau\}=0$.
Hence, the integrals in (\ref{B.13}) vanish as $\ve\to0$
by the Lebesgue--Riemann theorem.
The proof of convergence (\ref{B.11}) 
in the case when $d>1$ is similar and based on condition {\bf I4}.
\bo

%%%%%%%%%%%%%%%%%5


\begin{thebibliography}{99}

\bibitem{BPT} Boldrighini, C., Pellegrinotti, A., and Triolo, L.,
Convergence to stationary states for infinite harmonic systems,
{\it J. Stat. Phys.} {\bf 30} (1983), 123-155. 

\bibitem{DeMasi}
de Masi, A., Ianiro, N., Pellegrinotti, A., Presutti, E., A survey
of the hydrodynamical behavior of many-particle systems.
In: {\it Nonequilibrium phenomena. II. From stochastic to hydrodynamics.}
Lebowitz, J.L., Montroll, E.W. (eds.) pp. 123-294, Amsderdam: 
North-Holland, 1984.

\bibitem{DPST}
Dobrushin, R.L., Pellegrinotti, A., Suhov, Yu.M., and Triolo, L.,
One dimensional harmonic lattice caricature
of hydrodynamics, {\em J. Stat. Phys.} {\bf 43} (1986), 571-607.

\bibitem{DSS} Dobrushin, R.L., Sinai, Ya.G., and Sukhov, Yu.M., 
"Dynamical systems of statistical mechanics and kinetic equations. 
Chapter~10. Dynamical systems of statistical mechanics", 
{\it Dynamical systems}~--~2, Itogi Nauki i Tekhniki. Ser. Sovrem. Probl.
Mat. Fund. Napr., {\bf 2}, VINITI, Moscow, 1985, 235-284
(English transl., "Dynamical systems of statistical mechanics",
{\it Dynamical systems. II. Ergodic theory with applications 
to dynamical systems and statistical mechanics}, 
Encyclopaedia Math. Sci., vol.2, Springer-Verlag,
Berlin 1989, pp.207-278.)

%\bibitem{DKKS} Dudnikova, T.V., Komech, A.I., Kopylova, E.A., 
%and Suhov, Yu.M.,
% On convergence to  equilibrium distribution, I. The Klein-Gordon   
%equation with mixing, {\it Comm. Math. Phys.}  
%{\bf 225} (2002), no.1, 1-32.   

%\bibitem{DKRS} Dudnikova, T.V., Komech, A.I., Ratanov, N.E., and Suhov, Yu.M.,
%On convergence to   equilibrium distribution, II.
%The  wave  equation in odd dimensions, with mixing,  
% {\it  J. Stat. Phys.} {\bf 108} (2002), no.4, 1219-1253.  

\bibitem{DKS} Dudnikova, T.V., Komech, A.I., and Spohn, H.,   
On a  two-temperature problem for wave equation,   
 {\it Markov Processes and Related Fields} {\bf 8} (2002), 43-80.   

\bibitem{DK2}  Dudnikova, T., and Komech, A., 
On a two-temperature problem for the Klein-Gordon equation,
{\em Teor. Veroyatnost. i Primenen.} {\bf 50} (2005), no.4,
675-710 [in Russian] (English transl. 
{\em Theory Probab. Appl.} {\bf 50} (2006), no.4, 582-611).

\bibitem{DKS1}  Dudnikova, T.V., Komech, A.I., and Spohn, H., 
 On the convergence to statistical equilibrium for harmonic crystals,
 {\em J. Math. Phys.} {\bf 44} (2003), 2596-2620.

\bibitem{DKM}  Dudnikova, T.V., Komech, A.I., and Mauser, N.J.,
On two-temperature problem for harmonic crystals,
  {\em J. Stat. Phys.} {\bf 114} (2004), no.3/4, 1035-1083.

\bibitem{DS} Dudnikova, T.V., and Spohn, H., 
Local stationarity for lattice dynamics in the
harmonic approximation, {\em Markov Processes and Related Fields}
{\bf 12} (2006), no.4, 645-578. 
 
\bibitem{D08}  Dudnikova, T.V.,
On the asymptotical normality 
of statistical solutions for harmonic crystals in half-space,
{\it Russian J. Math. Phys.}  {\bf 15} (2008), no.4, 460-472.

\bibitem{IL} Ibragimov, I.A., and Linnik, Yu.V.,   
{\it Independent and Stationary Sequences of Random Variables} (1971),   
Ed. by J. F. C. Kingman,  Wolters-Noordhoff, Groningen.   

%\bibitem{LL}  Lanford III, O.E., and Lebowitz, J.L.,   
%{\it Time Evolution and Ergodic Properties of Harmonic Systems},  in:   
%{\it Dynamical Systems, Theory and Applications},   
% Lecture Notes in Physics {\bf 38} (1975), Springer-Verlag, Berlin.  

%\bibitem{M} A. Mielke,
%Macroscopic behavior of microscopic oscillations
%in harmonic lattices. February 2004.
%Preprint 118. {\em Analysis, Modeling and Simulation of Multiscape Problems}.
%Schwerpunktprogramm DFG.

\bibitem{P} Petrov, V.V., {\it Limit Theorems of Probability Theory},   
Clarendon Press, Oxford, 1995.   

\bibitem{Ros} Rosenblatt, M.A.,
A central limit theorem and a strong mixing condition,  
{\it Proc. Nat. Acad. Sci. U.S.A.} {\bf 42} (1956), no.1, 43-47.  

%\bibitem{SL} Spohn, H., and Lebowitz, J.L.,   
%Stationary non equilibrium states of infinite harmonic systems,   
%{\it Comm. Math. Phys.} {\bf 54} (1977), 97-120.    

\bibitem{Sp91} Spohn, H., {\em Large Scale Dynamics of Interacting
Particles}, Texts and Monographs in Physics, Springer
Verlag, Heidelberg, 1991.

\bibitem{Sp05} Spohn, H., The phonon Boltzmann equation, properties
and link to weakly anharmonic lattice dynamics, 
{\it J. Stat. Phys.} {\bf 124} (2006), no.2-4, 1041-1104.

\bibitem{VF} Vishik, M.I., and Fursikov,  A.V.,   
{\it Mathematical Problems of Statistical Hydromechanics}, 1988. 
Dordrecht: Kluwer Academic Publishers. 

%\bibitem{F} Fedoryuk,  M.V., The
%stationary phase method and pseudodifferential operators,
%{\em Russ. Math. Surveys} {\bf 26} (1971), 65-115.

\end{thebibliography}
\end{document}